\documentclass[12pt]{article}
\usepackage{amsfonts,amssymb,epsfig,amsmath}
\usepackage{color}

\usepackage[vcentermath]{youngtab}

\renewcommand{\baselinestretch}{1.2}

\begin{document}

\makeatletter \@addtoreset{equation}{section} \makeatother
\renewcommand{\theequation}{\thesection.\arabic{equation}}
\renewcommand{\thefootnote}{\alph{footnote}}

\begin{titlepage}

\begin{center}
\hfill {\tt KIAS-P11078}\\
\hfill {\tt SNUTP11-011}\\

\vspace{2cm}

{\Large\bf Vortices and 3 dimensional dualities}

\vspace{2cm}

\renewcommand{\thefootnote}{\alph{footnote}}

{\large Hee-Cheol Kim$^1$, Jungmin Kim$^2$, Seok Kim$^2$ and Kanghoon Lee$^3$}

\vspace{1cm}

\textit{$^1$School of Physics, Korea Institute for Advanced Study,
Seoul 130-722, Korea.}

\textit{$^2$Department of Physics and Astronomy \& Center for
Theoretical Physics,\\
Seoul National University, Seoul 151-747, Korea.}

\textit{$^3$Center for Quantum Spacetime, Sogang University, Seoul 121-742, Korea.}\\

\vspace{0.7cm}

E-mails: {\tt heecheol1@gmail.com, kjmint82@gmail.com, skim@phya.snu.ac.kr,
kanghoon@sogang.ac.kr }

\end{center}

\vspace{1.5cm}

\begin{abstract}

We study a supersymmetric partition function of topological vortices in 3d
$\mathcal{N}\!=\!4,3$ gauge theories on $\mathbb{R}^2\times S^1$, and use it to
explore Seiberg-like dualities with Fayet-Iliopoulos deformations. We provide
a detailed support of these dualities and also clarify the roles of vortices.
The $\mathcal{N}\!=\!4$ partition function confirms the proposed Seiberg duality and
suggests nontrivial extensions, presumably at novel IR fixed points with
enhanced symmetries. The $\mathcal{N}\!=\!3$ theory with nonzero Chern-Simons term also
has non-topological vortices in the partially broken phases, which are essential for the
Seiberg duality invariance of the spectrum. We use our partition function to confirm some
properties of non-topological vortices via Seiberg duality in a simple case.

\end{abstract}

\end{titlepage}

\renewcommand{\thefootnote}{\arabic{footnote}}

\setcounter{footnote}{0}

\renewcommand{\baselinestretch}{1}

\tableofcontents

\renewcommand{\baselinestretch}{1.2}

\section{Introduction}

Supersymmetric gauge theories in 3 dimensions have been studied extensively
in recent years with various motivations. Many 3d theories are related
to others by dualities which include strong-coupling physics. Various quantities were
studied to have a detailed understanding of these strongly coupled theories, often
relying on supersymmetry and localization methods. Especially, the superconformal
index \cite{Kim:2009wb} and the partition function on the 3-sphere \cite{Kapustin:2009kz}
found a wide range of applications. Although these quantities were originally calculated
for supersymmetric Chern-Simons-matter theories \cite{Schwarz:2004yj}, they are applicable
to a broader class of 3d theories. These partition functions were generalized and studied
in various extended frameworks \cite{Jafferis:2010un}.

In this paper, we study another quantity which contains interesting information
on the non-perturbative physics of 3d gauge theories. 3d gauge theories,
including CFT's with relevant deformations, in their Higgs phases can have vortex
solitons in their spectra. In supersymmetric gauge theories, BPS vortices often play
important roles in the dynamics of these theories \cite{Aharony:1997bx}.
We calculate and study an index which counts these BPS vortex particles.
The vortices we study are called topological vortices. In some theories, there are
also non-topological vortices in partially unbroken phases, which we only discuss
briefly. We use this partition function to investigate strong-coupling dualities
in 3 dimensions.

The vortex partition function (or the index) has been studied to certain extent
in the literature, and especially the partition function of 2 dimensional vortices
as instantons was investigated in \cite{Shadchin:2006yz,Dimofte:2010tz,Miyake:2011yr}.
See also \cite{Fujimori:2012ab} for a very recent work. The partition function of 3d
theories compactified on a circle was discussed as well. As instantons in Euclidean QFT
can be regarded as solitonic particles in
one higher dimension, the last vortex partition function has a natural
interpretation as an index counting BPS states.\footnote{A similar study for the index of 5
dimensional instanton particles are also done. In particular, the index of 5d maximal SYM
was recently studied \cite{Kim:2011mv} in the context of 6d $(2,0)$ theory
compactified on a circle.} However, see section 2.3 for a possible subtlety of the
index interpretation for the so-called semi-local vortices, due to their noncompact internal
moduli.

From the recent finding that squashed 3-sphere partition functions admit factorizations
to vortex partition functions \cite{Pasquetti:2011fj} in some theories,
the latter may perhaps be a more basic quantity than other 3d partition functions
\cite{Kim:2009wb,Kapustin:2009kz,Jafferis:2010un} known in the literature.
Also, the superconformal index which counts monopole operators is conceptually quite
similar to the vortex partition functions, as monopole operators are creating/annihilating
nonzero vortex charges. See also \cite{Dimofte:2011py} for comments on possible relations
of these partition functions in simple models.

In this paper, we restrict our study to 3d $\mathcal{N}\!=\!4$ and $3$ supersymmetric gauge
theories with $U(N)$ gauge group coupled to matter fields in fundamental representations.
Our partition function can be used to study various non-perturbative dualities of these
theories. We study the 3d Seiberg-like dualities similar to \cite{Aharony:1997gp}.
Seiberg duality is an IR duality, in which two different UV theories
flow to the same IR fixed point. Although it was originally found in 4d
$\mathcal{N}\!=\!1$ SQCD \cite{hep-th/9411149}, such dualities were also discovered and
studied in 3 dimensions. 3d Seiberg dualities were first discussed
in \cite{Aharony:1997gp}. Although they have some formal similarities to 4d Seiberg dualities,
physical implications of these dualities are not quite the same in different dimensions.
4d Seiberg duality can be regarded as an electromagnetic duality and also as a weak-strong
coupling duality \cite{hep-th/9411149}. Similar interpretation in 3d is lacking,
at least when there are no Chern-Simons term so that we do not have IR couplings.

Seiberg duality also exists after introducing a Fayet-Iliopoulos (FI) deformation
on both sides of the dual pair. Denoting by $\zeta$ the
FI parameter, there exist BPS vortex solitons whose masses are proportional to $\zeta$.
Considering the regime in which $\zeta$ is much smaller than the Yang-Mills coupling scale
$g_{YM}^2$, 3d Seiberg duality should map different types of `light' vortex
particles in the dual pair. We show that the spectra of the topological vortex particles
as seen by our partition function perfectly agree between the Seiberg-dual pairs, when they
exhaust all possible vortices (without non-topological vortices). This is the 3d version of
the 4d Seiberg duality map. While in the latter case glueballs, baryons and magnetic monopoles
in the confining phase map to the elementary quarks in the dual Higgs phase \cite{hep-th/9411149},
in 3d we naturally find that vortices map to dual vortices.

3d $\mathcal{N}\!=\!4$ Seiberg duality can be partly motivated by brane systems
\cite{Hanany:1996ie}. The $\mathcal{N}\!=\!4$ that we consider in this paper 
can be engineered by the D-brane/NS5-brane system shown in Fig \ref{n=4-brane}. 
By changing the positions of the two NS5-branes, which makes them cross
each other when $\zeta=0$, one obtains another 3d gauge theory with $U(N_f\!-\!N)$ gauge
group and $N_f$ flavors for $N_f\geq N$. \textit{Supposing that} both theories
flow to nontrivial IR fixed points, Seiberg duality asserts that the two IR fixed points are
the same. However, as pointed out in \cite{Gaiotto:2008ak,Kapustin:2010mh}, it
turns out that one of the two UV theories in the putative dual pair often does not flow to
an IR CFT, at least not in the `standard way' \cite{Gaiotto:2008ak} in which the
$SO(4)$ superconformal R-symmetry is the $SO(4)$ R-symmetry visible in UV. One way to
see this is to study the R-charges of BPS monopole operators, which we review in section 3.
Firstly, when $N_f<2N-1$,
there exist monopole operators whose R-charges are smaller than $\frac{1}{2}$.
If the theory flows to a CFT, the BPS bound demands that this R-charge be the scale dimension
of the operator, violating the unitarity bound. These theories were called `bad'
\cite{Gaiotto:2008ak,Kapustin:2010mh}. When $N_f=2N-1$, called `ugly' case, there exists
a monopole operator with R-charge $\frac{1}{2}$, saturating the unitarity bound in IR.
In this case, the modified version of the naive Seiberg duality is that the $U(N)$ theory
with $N_f=2N-1$ flavors is dual to the $U(N_f-N)=U(N-1)$ theory with $N_f=2N-1$ flavors times
a decoupled free theory of a (twisted) hypermultiplet. This has been recently tested from the
3-sphere partition function \cite{Kapustin:2009kz}. As the case with $N_f=2N$ is trivially self
Seiberg-dual, the pair containing the case with $N_f=2N-1$ was the only nontrivial dual pair \cite{Gaiotto:2008ak,Kapustin:2010mh}.

Our vortex partition function confirms this duality at $N_f=2N-1$: namely, the
partition function agrees with that of the $U(N-1)$ theory with $N_f=2N-1$ flavors times
the vortex partition function of the $N=N_f=1$ theory (the Abrikosov-Nielsen-Olesen vortex).
As the last vortices are free, it agrees with the above argument that the free hypermultiplet
sector exists. Also, the monopole operator mentioned in the previous paragraph with dimension
$\frac{1}{2}$ is nothing but the vortex-creating operator, making it natural to identify
the above free hypermultiplet as the vortex supermultiplet. The test can be made at each
vortex number $k=1,2,3,\cdots$, which makes the confirmation highly nontrivial. Moreover,
our vortex partition function suggests that the general putative dual pair for any $N_f\geq N$
could be actually Seiberg-dual to each other, also with a modification by adding a factorized
sector. This may be suggesting a broader class of IR fixed points than those identified in
\cite{Gaiotto:2008ak}. See section 3.1 for the details.

Seiberg dualities with $\mathcal{N}\leq 3$ supersymmetry were also studied quite
extensively in recent years, after they were discovered in Chern-Simons-matter
theories \cite{Giveon:2008zn,Niarchos:2008jb,Niarchos:2009aa}. For instance,
\cite{Kapustin:2010mh,Kapustin:2011gh,Willett:2011gp,Bashkirov:2011vy,Hwang:2011qt,Hwang:2011ht}
studied various 3d Seiberg dualities using the 3-sphere partition function and the
superconformal index. Other studies on the Chern-Simons Seiberg dualities include
\cite{Amariti:2009rb}. We study the vortex partition function of the above $\mathcal{N}\!=\!4$
theory, deformed by an $\mathcal{N}=3$ Chern-Simons term. With nonzero FI
deformation, the vacuum structure becomes more complicated than that for the gauge theory
with zero Chern-Simons term, as one also finds partially Higgsed phases. The Seiberg
duality maps a branch of vacuum to another definite branch in the dual theory.
Due to the presence of partially unbroken Chern-Simons gauge symmetry, it turns out that
the study of non-topological vortices is also crucial for the Seiberg duality invariance
of the vortex spectrum. From our index for topological vortices, we study aspects of
the Seiberg-dual non-topological vortices. In some simple cases, our topological vortex
index confirms nontrivial properties of non-topological vortices suggested in the literature
via Seiberg duality. Namely, we show that the vorticity and angular momentum of non-topological
vortices in the Chern-Simons-matter theory with $N=N_f=1$ satisfy a bound required by
a tensionless domain wall picture of \cite{Kim:2006ee}, via topological vortex calculation
of the Seiberg-dual theory.

The remaining part of this paper is organized as follows. In section 2, we explain
$\mathcal{N}\!=\!4,3$ field theories, BPS topological vortices, and then
derive the vortex partition functions (or indices). In section 3, we show that these
partition functions nontrivially confirm the known Seiberg dualities of
some $\mathcal{N}\!=\!4$ theories with FI deformations. We then suggest a wide extension of
this duality, presumably at new kinds of IR fixed points. We also study $\mathcal{N}\!=\!3$
Seiberg dualities from vortices. Section 4 concludes with discussions.
Appendices A, B, C explain the structure of vortex quantum mechanics and also
a derivation of the vortex partition function.

\section{Vortex partition functions of 3d gauge theories}

\subsection{Supersymmetric gauge theories and vortices}

Let us first consider the $\mathcal{N}\!=\!4$ $U(N)$ gauge theory with $N_f$ fundamental
hypermultiplets. Its bosonic global symmetry is $SO(2,1)$ coming from spacetime,
$SO(4)=SU(2)_L\times SU(2)_R$ R-symmetry which rotates the supercharges as a vector,
and an $SU(N_f)$ flavor symmetry. It has a vector supermultiplet consisting of the gauge field
$A_\mu$, gaugino $\lambda^{a\dot{b}}_\alpha$ (where $a,\dot{b}$ are $SU(2)_L$ and $SU(2)_R$ doublet indices, respectively, and $\alpha$ is for $SO(2,1)$ spinors), and three scalars $\phi^I$
($I=1,2,3$ for $SU(2)_L$ triplet). The hypermultiplets consist of
$N_f$ pairs of complex scalars $q_{\dot{a}}^i$ in the fundamental representation of $U(N)$,
where $i=1,2,\cdots, N_f$, and superpartner fermions $\psi^{ai}_\alpha$. The supercharges
$Q^{a\dot{b}}_\alpha$ are taken to be Majorana spinors involving $SU(2)_L\times SU(2)_R$
conjugations. The bosonic part of the action is given by
\begin{equation}\label{bosonic-N=4}
  \mathcal{L}_{\rm bos}=\frac{1}{g_{YM}^2}{\rm tr}\left[-\frac{1}{4}F_{\mu\nu}F^{\mu\nu}
  -\frac{1}{2}D_\mu\phi^I D^\mu\phi^I+\frac{1}{4}[\phi^I,\phi^J]^2\right]
  -D^\mu q^\dag_{i\dot{a}}D_\mu q^{i\dot{a}}
  -q^\dag_{i\dot{a}}\phi^I\phi^Iq^{i\dot{a}}-\frac{1}{2g_{YM}^2}D^AD^A
\end{equation}
where $A=1,2,3$ is the triplet index for $SU(2)_R$, and
$g_{YM}^{-2}D^A=q^{i\dot{a}}(\tau^A)_{\dot{a}}^{\ \dot{b}}q^\dag_{i\dot{b}}$ with three
Pauli matrices $\tau^A$. The moduli space of this theory has two parts.
The classical Coulomb branch is obtained by taking $\phi^I$ to be nonzero and all diagonal,
while all hypermultiplet scalars are set to zero. The Higgs branch is obtained by taking
$\phi^I=0$, while nonzero $q^{i\dot{a}}$ satisfy the condition $D^A\!=\!0$. The real
dimension of the Higgs branch moduli space (modded out by the action of gauge transformation)
is $4N(N_f-N)$. As we shall be
mainly interested in the Higgs branch which supports vortex solitons, and also due to the
motivation of studying Seiberg duality, we shall restrict our studies to the theories
satisfying $N\leq N_f$. The Coulomb and Higgs branches meet at least at a point in which
all fields are set to zero. They may meet more nontrivially in the presence of the vacua
with partially unbroken gauge symmetry when $N_f\leq 2N-1$ \cite{Gaiotto:2008ak}.

One can also introduce Fayet-Iliopoulos deformations for the overall $U(1)$ part
of $U(N)$, which leaves the form of the (bosonic) action as (\ref{bosonic-N=4}) but
changes the D-term fields $D^A$ to
\begin{equation}
  g_{YM}^{-2}D^A=q^{i\dot{a}}(\tau^A)_{\dot{a}}^{\ \dot{b}}q^\dag_{i\dot{b}}-\zeta^A\ ,
\end{equation}
with three real constants $\zeta^A$. Nonzero FI parameters break $SU(2)_R$
to $U(1)$. Without losing generality, we can take $\zeta\equiv\zeta^3>0$
and other two to be zero. It will also be convenient to call $q^i\equiv q^{i\dot{1}}$
and $\tilde{q}_i\equiv q^\dag_{i\dot{2}}$. The vacuum condition
$D^A=0$ can be written as
\begin{equation}\label{higgs}
  q^i\tilde{q}_i=0\ ,\ \ q^iq_i^\dag-\tilde{q}^{i\dag}\tilde{q}_i=\zeta\ .
\end{equation}
The hypermultiplet scalar should be nonzero and totally break $U(N)$ gauge symmetry,
lifting the Coulomb branch. With $\zeta>0$, a subspace of the Higgs branch moduli space which
will be useful later is obtained by setting $\tilde{q}_i=0$. The second equation
of (\ref{higgs}) is then solved by picking a $U(N)$ subgroup of $SU(N_f)$, and taking
\begin{equation}
  q=\left(\sqrt{\zeta}\ {\bf 1}_{N\times N}\ |\ {\bf 0}_{N\times(N_f\!-\!N)}\right)\ ,
\end{equation}
where we view $q^i$ as an $N\times N_f$ rectangular matrix. The possible embeddings of
$U(N)$ yields a vacuum moduli subspace given by the Grassmannian
$\frac{SU(N_f)}{S[U(N)\times U(N_f-N)]}$. At any point,
$S[U(N)\times U(N_f-N)]=\frac{U(N)\times U(N_f-N)}{U(1)}$ global symmetry remains unbroken.

With nonzero $\zeta$ ($>\!0$), there exist BPS vortex solitons on the above subspace
given by $\tilde{q}_i=0$. The BPS equations
can be obtained either from supersymmetry transformations or by complete-squaring
the bosonic Hamiltonian \cite{Hanany:2003hp}, which are
\begin{equation}
  F_{12}=g_{YM}^2(q^iq_i^\dag-\zeta)\ ,\ \ (D_1-iD_2)q^i=0\ ,\ \
  k\equiv-\frac{1}{2\pi}\int d^2x\ {\rm tr}F_{12}\ (\in\mathbb{Z})\ >0\ .
\end{equation}
We have chosen to study vortices with $k>0$, rather than
anti-vortices. The BPS mass of the vortices is given by $2\pi\zeta k$.
There is a moduli space of the solution, with real dimension
$2kN_f$ \cite{Hanany:2003hp}. These vortices preserve $4$ real supersymmetries
$Q^{a\dot{1}}_{-}$ and $Q^{a\dot{2}}_+\sim \epsilon^{ab}(Q^{{b}\dot{1}}_-)^\dag$
among the full $\mathcal{N}=4$ supercharges $Q^{a\dot{b}}_\alpha$, where $\pm$ denote
$SO(2,1)$ spinor components in the eigenspinor basis of $\gamma^0$.
The vortex quantum mechanics model which we introduce
later explicitly preserves $Q^{a\dot{1}}_-$, which shall be written as $Q^a$.

The nature of topological vortices depends on whether $N_f\!=\!N$ or $N_f\!>\!N$.
When $N_f\!=\!N$, the
$2kN$ dimensional vortex moduli space consists of $2k$ translation zero modes of $k$
vortices and $2k(N-1)$ internal zero modes. The $N-1$ complex zero mode per vortex
can be understood as the embedding moduli of $U(1)$ Abrikosov-Nielsen-Olesen (ANO) vortex
into $U(N)$. Namely, the internal moduli of a single vortex is
$\mathbb{CP}^{N-1}=\frac{U(N)}{U(N-1)\times U(1)}$.
When $N_f>N$, $2k(N_f\!-\!N)$ extra internal zero modes exist.
There are noncompact directions from these extra modes, as vortices
can now come with size moduli. These vortices are called semi-local vortices.

The low energy dynamics of these vortices can be studied in various ways.
It can be studied by a D-brane realization of the QFT and vortices \cite{Hanany:2003hp},
as we shall review shortly. Also, one can perform a careful moduli space approximation
in the field theory context, which has been done in \cite{Shifman:2006kd,Eto:2006uw}.
It turns out that some of the dynamical degrees kept in the naive D-brane considerations
\cite{Hanany:2003hp} originates from non-normalizable zero modes
\cite{Shifman:2006kd,Eto:2006uw} from the field theory viewpoint. More concretely,
supposing that we introduce an IR cut-off regularization of length scale $L$,
it was shown that the mechanical kinetic terms for the last modes pick up a
factor proportional to $\log L$ \cite{Shifman:2006kd,Eto:2006uw}. After carefully
redefining variables in a way that IR divergence does not appear, it was shown
in the single vortex sector that the K\"{a}hler potential for the quantum mechanical
sigma model differs from that derived from the D-brane approach \cite{Shifman:2006kd}.

Let us explain the difference in some detail and clarify our viewpoint on
the index calculation. Our claim is that the index
will be the same no matter which vortex quantum mechanics is used, as the index is
insensitive to various continuous parameters of the theory. If one could find a continuous
supersymmetric  deformation between the string-inspired model of \cite{Hanany:2003hp} and
more rigorously derived field theory models, this would prove our claim. Actually at $k\!=\!1$,
the two K\"{a}hler potentials of \cite{Hanany:2003hp} and \cite{Shifman:2006kd} can be written as
\begin{equation}
  K_{\rm HT}=\sqrt{r^2+4r|\zeta|^2}-r\log\left(r+\sqrt{r^2+4r|\zeta|^2}\right)
  +r\log(1+|z_i|^2)\ ,\ \ K_{\rm SVY}=|\zeta|^2+r\log(1+|z_i|^2)
\end{equation}
with $|\zeta|^2\equiv(1+|z_i|^2)|z_p|^2$,
where the summations over $i$ range in $1,2,\cdots,N\!-\!1$, and those over $p$ range
in $N,\cdots N_f\!-\!1$. Deforming the former K\"{a}hler potential to the latter one
in a continuous way will prove that there is a supersymmetric deformation between the
two. Of course there is an issue on the non-compact region. As we shall illustrate with
detailed calculations in the appendices, our index can be completely determined from
the information of the moduli space near the region where $z_p=0$, at which the vortex
sizes are minimal. So we can ignore any possible difference in the asymptotic behaviors
of the two metrics. By inserting $\epsilon|\zeta|^2$ to all $|\zeta|^2$ in $K_{\rm HT}$,
and also multiplying $\epsilon^{-1}$ to the first two terms of $K_{\rm HT}$, one obtains
a 1-parameter deformation of the Kahler potential. By taking the $\epsilon\rightarrow 0$ limit,
one finds that $K_{\rm HT}$ reduces to the exact field theory result $K_{\rm SVY}$.

Although the above kind of comparison can be made only when the moduli space metric is
explicitly known, we expect the same phenomenon to appear for multi-vortices. This is because
our index is only sensitive to the region near minimal size semi-local vortices, and all
concrete studies from QFT \cite{Shifman:2006kd,Eto:2006uw} suggest that the difference between
the two approaches will be absent in this region. In particular, the second reference in
\cite{Eto:2006uw} discusses this point for some multi-vortex configurations.
We also mention that \cite{Shifman:2006kd} finds certain BPS spectrum of the two models agree
with each other. In the rest of this paper, we shall be working with the models like
\cite{Hanany:2003hp} derived from the naive D-brane pictures, to derive the index.

\begin{figure}[t!]
  \begin{center}
    \includegraphics[width=17cm]{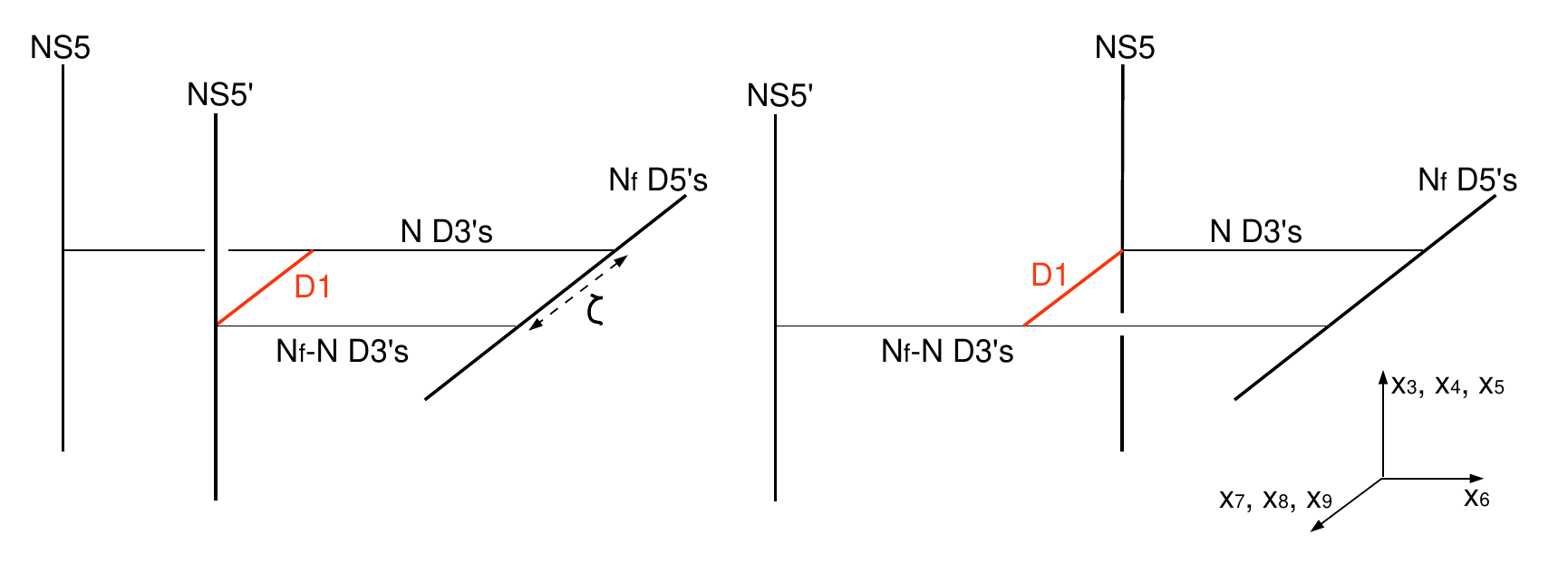}
\caption{The brane construction of $\mathcal{N}\!=\!4$ Seiberg-dual pairs with nonzero
FI parameter $\zeta$. The red lines denote D1-brane vortices.}\label{n=4-brane}
  \end{center}
\end{figure}
One can engineer the above gauge theories and vortices from branes in
type IIB string theory, as shown in Fig \ref{n=4-brane}. The D3-branes,
NS5-branes, D5-branes are along $0126$, $012345$, $012789$ directions,
respectively. When $\zeta=0$, $N$ D3-brane segments connect two NS5-branes
and can move in the $345$ direction along the NS5-brane worldvolume. This forms
the Coulomb branch, with $\mathbb{R}^3$ showing the $SU(2)_L$ symmetry. Extra $N_f$
D3-branes connect the NS5-brane on the right side (NS5$^\prime$) and the $N_f$ D5-branes.
The open strings connecting the two sets of D3-branes provide the fundamental
hypermultiplet matters. The Seiberg duality that we shall explore in this
paper corresponds to moving the two NS5-branes across each other. By the brane
creation effect \cite{Hanany:1996ie}, the number of D3-branes
between the two NS5-branes after this crossing is $N_f\!-\!N$.

Turning on nonzero FI term corresponds to moving one NS5-brane along its transverse
$789$ directions, parametrized by the $SU(2)_R$ triplet $\zeta^A$. The $N$ D3-branes
cannot finish on NS5$^\prime$-brane preserving supersymmetry. So one has
to combine them with $N$ of the $N_f$ flavor D3-branes, as shown on the left side of
Fig \ref{n=4-brane}. The remaining $N_f\!-\!N$ D3-branes connect the NS5$^\prime$-brane
and the D5-branes. Fig \ref{n=4-brane} shows the two brane configurations after we move
two NS5-branes across them in the $6$ direction. In both cases, there exist BPS D1-branes
(as shown by the red segment), corresponding to the BPS vortices.

It will also be helpful to understand the supersymmetry preserved by
these branes. The NS5-, D5-, and D3-branes preserve supersymmetries
which satisfy definite projection conditions for $(\sigma_3)\otimes\Gamma^{012345}$,
$(\sigma_1)\otimes\Gamma^{012789}$ and $(i\sigma_2)\otimes\Gamma^{0126}$
\cite{Bergshoeff:1996tu,Imamura:1998gk}, where $\Gamma^{0123456789}=+1$ from type IIB
chirality. All three projectors commute. From
\begin{equation}
  (\sigma_3)\otimes\Gamma^{012345}=-\left[(\sigma_1)\otimes\Gamma^{012789}\right]
  \cdot\left[(i\sigma_2)\otimes\Gamma^{0126}\right]\ ,
\end{equation}
two projection conditions imply the third. So one finds a $1/4$-BPS
configuration, preserving $8$ real or 3d $\mathcal{N}\!=\!4$ SUSY. More concretely,
the $8$ SUSY may be obtained as follows. Taking the $6$ commuting
matrices, $A=(i\sigma_2)\otimes\Gamma^0$, $B=\Gamma^{12}$; $C=\Gamma^{34}$,
$D=(\sigma_1)\otimes\Gamma^5$; $E=\Gamma^{78}$, $F=(\sigma_3)\otimes\Gamma^9$, one
can write $\Gamma^6=-ABCDEF$ and also write the $3$ projectors as
\begin{equation}
  \Gamma_{NS5}=+ABCD\ ,\ \ \Gamma_{D5}=-ABEF\ ,\ \ \Gamma_{D3}=AB\Gamma^6=+CDEF\ ,
\end{equation}
respectively. The $32$ real components of the type IIB spinor can be obtained by starting
from $64$ dimensional real spinor (with $32$ components from 10d and $2$ components from
$SL(2,\mathbb{R})$ Pauli matrices), and subjecting them to the chirality condition. However,
the matrices $A,D,F$ do not commute with the chirality operator $\Gamma^{11}$. So to
obtain the eigenspinors of the BPS projections using $A,B,C,D,E,F$ eigenstates,
one would always have to make a linear combination of different eigenstates of $A,D,F$
at the final stage, to make them eigenstates of $\Gamma^{11}$.
Supposing that 1st/3rd projectors for NS5/D3-branes come with $+1$ eigenvalues,
one has $AB=\pm i$, $CD=\mp i$, $EF=\pm i$, where the $\pm$ signs are correlated.
The possible signs of the eigenvalues and eigenvectors of $(A,B,C,D,E,F)$ come in
$8$ cases $\Psi_{s_1,s_2,s_3}\sim(s_1,s_1;s_2,-s_2;s_3,s_3)$ for the upper signs,
and $8$ cases $\Upsilon_{s_1,s_2,s_3}\sim(s_1,-s_1;s_2,s_2;s_3,-s_3)$ for the lower signs,
where $s_1,s_2,s_3$ are independent $\pm$ signs. Since the $A,D,F$ do not commute with
$\Gamma^{11}$ but rather anticommute, we should mutiply $\frac{1+\Gamma^{11}}{2}$ to the
spinors to get 10d chiral spinors. We can define $\Gamma^{11}\Psi_{s_1,s_2,s_3}\equiv\Upsilon_{-s_1,s_2,s_3}$.
The chirality projection only keeps the combination
$\Psi_{s_1,s_2,s_3}+\Upsilon_{-s_1,s_2,s_3}$, and we finally have $8$ or 3d
$\mathcal{N}=4$ SUSY.

The supersymmetry for D1-brane vortex is given by the projector
$(\sigma_1)\otimes\Gamma^{09}=-AF$ \cite{Imamura:1998gk}, supposing that the
FI parameter is separating two NS5-branes along the $9$ direction. This again
commutes with the remaining two
projections, and makes the vortex preserve $4$ real SUSY. More concretely, let us
assume that $AF$ has $+1$ eigenvalue. In the above two classes of $\pm$ sign,
we can take either $A=\pm 1$, $F=\pm 1$ before $\Gamma^{11}$ projection, yielding
$(s_1,s_1;s_2,-s_2;s_1,s_1)$ or $(s_1,s_1;s_2,s_2;-s_1,s_1)$. Thus, one has $8$ spinors
before projection, and after $\Gamma^{11}$ projection one obtains $4$ SUSY. As the two
NS5-branes cross, the D1-brane in the $U(N)$ theory and $U(N_f-N)$ theory appears as
vortex/anti-vortex, respectively, depending on whether the D-string starts or ends
on the brane on which the 3d gauge theory lives. We should thus compare the vortex
and anti-vortex spectra of the two theories.

One can also study vortices in the $\mathcal{N}=3$ theory with an FI deformation.
We first review the $\mathcal{N}=3$ theory with FI term and its vacua. One obtains
the $\mathcal{N}=3$ Yang-Mills Chern-Simons gauge theory by adding a Chern-Simons
term to the above $\mathcal{N}=4$ theory. Keeping the three D-term fields $D^A$ off-shell,
one adds to the action the following
Chern-Simons term
\begin{equation}
  \frac{\kappa}{4\pi}\int{\rm tr}\left(\epsilon^{\mu\nu\rho}(A_\mu\partial_\nu A_\rho
  -\frac{2i}{3}A_\mu A_\nu A_\rho)-2\phi^A D^A\right)+{\rm fermions}\ .
\end{equation}
The $SU(2)_L\times SU(2)_R$ R-symmetry of the $\mathcal{N}=4$ theory is reduced
to the diagonal $SU(2)$. So we no longer distinguish the $I$ and $A$ triplet indices of
two $SU(2)$'s, or the dotted/undotted doublet indices. By integrating out $D^A$,
one obtains the bosonic potential
\begin{equation}
  -\frac{1}{4g_{YM}^2}[\phi^A,\phi^B]^2+\frac{1}{2g_{YM}^2}D^AD^A\ ,\ \
  g_{YM}^{-2}D^A=q^{i\dot{a}}(\tau^A)_{\dot{a}}^{\ \dot{b}}q^\dag_{i\dot{b}}-\zeta^A
  -\frac{\kappa}{2\pi}\phi^A\ .
\end{equation}

The classical supersymmetric vacuum solutions can be found from the above
bosonic potential. One first finds that the Coulomb branch is lifted.
This is because $\phi^A$ acquires nonzero mass either from a superpartner
of the Chern-Simons term, or by the Higgs mechanism.
With $\zeta\neq 0$, one finds many partially Higgsed branches.
For simplicity, we shall only consider the subspace in which $\tilde{q}_i=0$
which admits topological BPS vortices. The classical supersymmetric vacuum
with $\tilde{q}_i=0$ can be obtained from the following equations,
\begin{equation}
  qq^\dag-\zeta-\frac{\kappa}{2\pi}\sigma=0\ ,\ \ \sigma q=0\ ,
\end{equation}
where we have set $\phi^1=\phi^2=0$ (with $\sigma\equiv\phi^3$) as they will also be
zero when the FI term is along the third component $\zeta\equiv \zeta^3$ only. The simplest
solution is obtained by setting $\sigma=0$. Then the condition $qq^\dag=\zeta$ simply yields the
$\frac{U(N_f)}{U(N)\times U(N_f-N)}$ moduli space. More generally \cite{arXiv:0805.0602},
we can take an $n\times n$ block of the $N\times N$ matrix $\sigma$
to be nonzero. Then from the second condition, $q$ has to sit in the $(N-n)\times N_f$
block orthogonal to $\sigma$. From the first equation, $qq^\dag$ and $-\frac{\kappa}{2\pi}\sigma$
are rank $N-n$ and $n$ matrices, respectively, which are acting on mutually orthogonal
subspaces. Thus the solution of the first equation is
\begin{equation}
  \sigma=-\frac{2\pi\zeta}{\kappa}{\bf 1}_{n\times n}\ ,\ \
  qq^\dag=\zeta{\bf 1}_{(N-n)\times (N-n)}\ \rightarrow\ \
  q\in\frac{U(N_f)}{U(N-n)\times U(N_f-N+n)}\ .
\end{equation}
The $U(N)_\kappa$ gauge symmetry is broken by the above vacuum to $U(n)_\kappa$.
From the quantum dynamics of this $U(n)_\kappa$ Chern-Simons gauge theory, supersymmetric
vacua exist only when $0\leq n\leq \kappa$ \cite{Witten:1999ds}.
Therefore, there are $\min(\kappa,N)+1$ branches of partially Higgsed
supersymmetric vacua, labeled by $n$ in the range $0\leq n\leq\min(\kappa,N)$.

\begin{figure}[t!]
  \begin{center}
    \includegraphics[width=17cm]{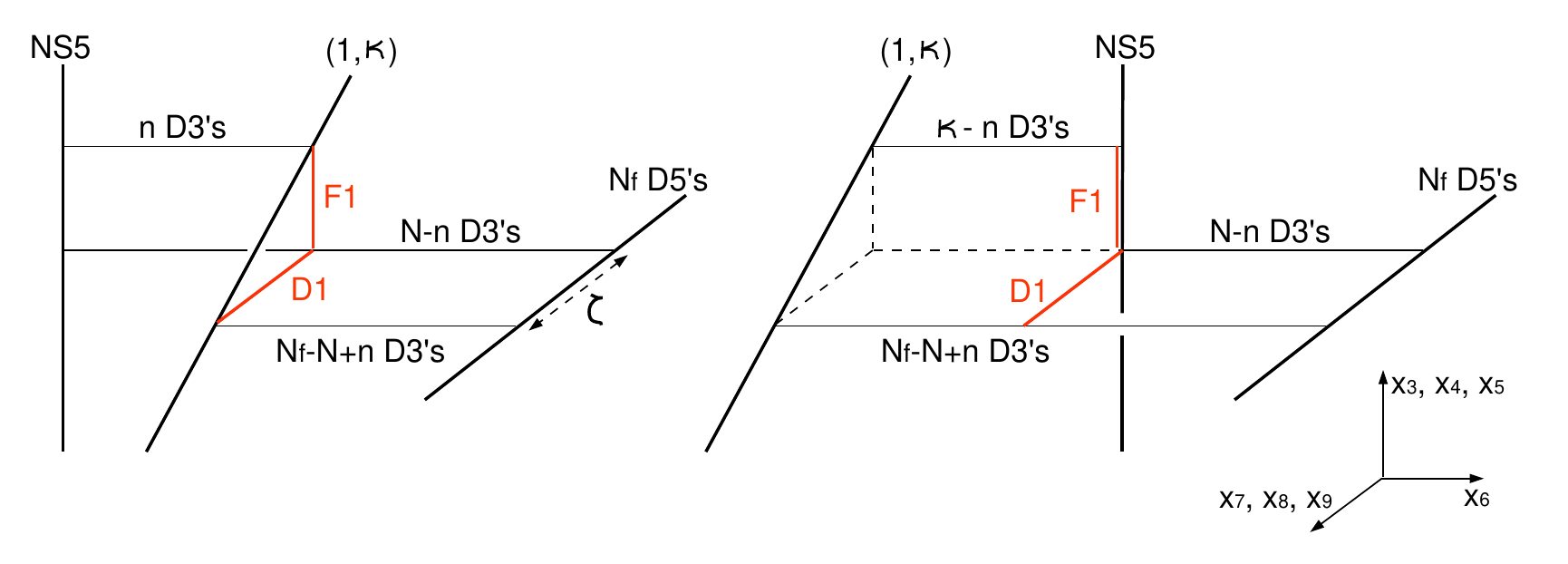}
\caption{The brane construction of $\mathcal{N}\!=\!3$ Seiberg-dual pairs and vortices
with unbroken $U(n)_\kappa$ or $U(k-n)_{-\kappa}$ gauge symmetry.}\label{n=3-brane}
  \end{center}
\end{figure}
It is helpful to consider all these aspects from the brane construction,
as shown in Fig \ref{n=3-brane}. The situation is similar to the $\mathcal{N}=4$
brane configuration, but to induce nonzero Chern-simons term, one changes the
second NS5-brane to an $(1,\kappa)$ 5-brane tilted in the $345$ and $789$ direction.
Namely, apart from the $012$ direction, the worldvolume of the $(1,\kappa)$-brane
has to be aligned along $x^3+\tan\theta x^7$, $x^4+\tan\theta x^8$, $x^5+\tan\theta x^9$
directions with $\tan\theta=\kappa g_s$, where $g_s$ is the type IIB coupling (at zero RR
0-form, which we assume for simplicity). Putting nonzero FI term $\zeta^A$ again corresponds to
moving the $(1,\kappa)$ brane relative to the NS5-brane in the $789$ direction. When $\zeta=0$,
again there are $N$ D3-branes connectiong NS5- and $(1,\kappa)$-branes, and also $N_f$ D3-branes
connecting $(1,\kappa)$- and D5-branes. When $\zeta\neq 0$, there are many possible deformations
of this D3-brane configurations, corresponding to various partially Higgsed phases.
On the left side of Fig \ref{n=3-brane}, there can be some fraction of $N$ D3-branes which
can connect NS5- and $(1,\kappa)$-branes even after FI deformation. We take $n$ D3-branes to do so.
The remaining $N\!-\!n$ of them should combine with the flavor D3-branes as shown in the figure,
whose gauge symmetry is spontaneously broken. The brane
configuration maps to the partially Higgsed branch with unbroken $U(n)_\kappa$ gauge symmetry.
The proposed Seiberg duality \cite{Giveon:2008zn} is obtained by moving NS5-
and $(1,\kappa)$-brane across each other. One then obtains a $U(N_f-N+|\kappa|)$ theory
coupled to $N_f$ fundamental hypermultiplets, at Chern-Simons level $-\kappa$. With brane
creations \cite{Hanany:1996ie}, the vacuum on the left side of Fig \ref{n=3-brane} maps
to the branch in the dual theory with unbroken $U(\kappa-n)_{-\kappa}$ gauge symmetry,
on the right side of the figure.

The BPS vortices in the $\mathcal{N}=3$ theory appear in many different ways.
We first consider them in the brane picture.
Firstly, there can be D1-branes connecting the $N-n$ D3-branes, corresponding
to the broken $U(N-n)$ gauge symmetry, and the $N_f-N+n$ D3-branes
corresponding to the remaining flavor branes. See the red horizontal
line on the left side of Fig \ref{n=3-brane}. As the D1-brane ends on the $N-n$ D3-brane
for broken gauge symmetry, they would correspond to topological vortices, similar to the
vortices in the $\mathcal{N}=4$ theories.
Actually, there exist 4d states given by this segment of D1-brane freely moving along
the D3-branes, behaving as monopoles in the decoupled 4d gauge theory. So the D1-branes
would be visible in the 3d theory as vortices only when they are marginally bound to the
5-brane, as shown in the figure.

There can also be vertical massive fundamental strings connecting the $n$ D3-branes
(with unbroken $U(n)_\kappa$ symmetry) and other D3-branes. If the FI parameter deformation
is made in the $9$ direction, the string is stretched in the $5$ direction. We shall shortly
show that this configuration preserves same SUSY as the D1-brane vortices. Also, as
$U(n)_\kappa$ gauge symmetry is unbroken, the overall $U(1)$ Noether charge (electric charge)
induces nonzero $\int{\rm tr}_{U(N)}F_{12}$ vorticity via the Gauss' law with $U(n)_\kappa$
Chern-Simons term. From the field theory perspective, these vortices are often called
non-topological vortices.

Also, there can be strings made of one D1 and $\kappa$ F1's, which vertically end on
the $(1,\kappa)$ 5-brane and $N\!-\!n$ D3-branes. This configuration preserves the same
SUSY as the above two types of vortices.
One can also show that the energy of this string is exactly the same as the D1-brane vortex
of first type, i.e. $2\pi\zeta$, by calculating the length and tension of the string.
It seems that our topological vortex index should be counting these configurations as well.

Although we follow \cite{Hanany:2003hp} to consider the $\mathcal{N}\!=\!3$ version
of their brane configuration given by Fig \ref{n=3-brane}, it is often clearer to move
$N_f$ D5's along $x^6$ to have it between the other two 5-branes \cite{Giveon:2008zn}.
See section 3.2 for more explanations.

It is easy to check the supersymmetry of these brane configurations. The NS5-, D5-,
$(1,\kappa)$- and the D3-branes require the projection conditions for
$(\sigma_3)\otimes\Gamma^{012345}$, $(\sigma_1)\otimes\Gamma^{012789}$,
$(c_\theta\sigma_3+s_\theta\sigma_1)\otimes\Gamma^{012(c_\theta 3+s_\theta 7)
(c_\theta 4+s_\theta 8)(c_\theta 5+s_\theta 9)}$, $(i\sigma_2)\otimes\Gamma^{0126}$.
To study the common eigenstates of these projectors, we again express all the
projections in terms of the six commuting projectors $A,B,C,D,E,F$. The eigenstates
of 3 projectors which are inherited from the $\mathcal{N}=4$ theory can again be solved
in terms of $8$ spinors $\Psi_{s_1,s_2,s_3}=(s_1,s_1;s_2,-s_2;s_3,s_3)$ and $8$ other spinors $\Upsilon_{s_1,s_2,s_3}=(s_1,-s_1;s_2,s_2;s_3,-s_3)$ before chirality projection.
The projection for the $(1,\kappa)$-brane is given by
\begin{equation}\label{1k-projection}
  AB\left(c_\theta^2D-s_\theta^2F-is_\theta c_\theta\sigma_2(D+F)\right)\left(c_\theta^2 C+s_\theta^2 E+s_\theta c_\theta\Gamma^{38}(1+CE)\right)\ .
\end{equation}
One way for this projector to have $+1$ eigenvalue is to have $DF=-1$ (real$^2$),
$CE=-1$ (imaginary$^2$) so that both parentheses in the projector yield $\pm 1$,
independent of $\theta$. In this case, one obtains from the above $16$ spinors the
following $8$ cases: $\Psi_{s_1,s_2,s_2}=(s_1,s_1;s_2,-s_2;s_2,s_2)$ or
$\Upsilon_{s_1,s_2,s_2}=(s_1,-s_1;s_2,s_2;s_2,-s_2)$. One also has to demand that the
projection for the $(1,\kappa)$-brane comes with a definite sign. $is_2$ is the last factor
including $C,E$, and $\mp s_2$ is the second factor including $D,F$, where $\mp$ is for
the $\Psi$/$\Upsilon$ cases. So the 2nd times 3rd factor becomes $\mp i$.
Since $AB$ in the 1st/2nd case is $\pm i$, this cancels with the $\mp i$ to always yield
$+1$. So we have $8$ components of spinors before chirality projection. The chirality
projection demands the combination $\Psi_{s_1,s_2,s_2}+\Upsilon_{-s_1,s_2,s_3}$, leaving
$4$ SUSY.

There is a different way of having (\ref{1k-projection}) satisfied. Rather than having
the second/third parenthesis to be separately $\theta$ independent numbers, the two
factors can yield $\theta$ dependent expression which cancel each other. So we start by
assigning $D=F=s$, which would yield $c_\theta^2D-s_\theta^2F-is_\theta c_\theta
\sigma_2(D+F)=se^{-2i\theta\sigma_2}$. Assigning definite eigenvalues for $D,F$ is
possible as the third factor does not change their eigenvalues. $\sigma_2$ operator
changes the eigenvalues of $D,F$ but leaves all other eigenvalues unchanged.
Now take $C=-E=is^\prime$. Then the last factor becomes $is^\prime e^{2\theta\Gamma^{38}}$.
The matrix $\Gamma^{38}$ changes the $C,E$ eigenvalues while leaving all the other
eigenvalues unchanged. The matrix $e^{-2i\theta\sigma_2}e^{2\theta\Gamma^{38}}$ can be
diagonalized by suitably mixing two states in $\Psi,\Upsilon$
with different signs $s_2$, $s_3$. Since we are restricted to the sector $D=F$, $C=-E$,
we only consider $\Psi_{s_1,s_2,-s_2}$ and $\Upsilon_{s_1,s_2,-s_2}$. The matrix  $e^{-2i\theta\sigma_2}e^{2\theta\Gamma^{38}}$ is expanded as
\begin{equation}
  \cos^2 2\theta+\sin^2 2\theta(-i\sigma_2\Gamma^{38})
  +\sin 2\theta \cos 2\theta\left(\Gamma^{38}-i\sigma_2\right)\ .
\end{equation}
The last linear terms are taking states out of the subspace which satisfies
the NS5-, D5-, D3-brane projections. So these terms should vanish by canceling
with each other. This freezes the linear combination of $\Psi_{s,+,-}$ and
$\Psi_{s,-,+}$, and also that of $\Upsilon_{s,+,-}$ and $\Upsilon_{s,-,+}$.
The remaining $-i\sigma_2\Gamma^{38}$ is also diagonalized then, with eigenvalue
$+1$, making the whole projection to be $+1$.
We thus have two $\Psi$ type states with two values for $s_1$,
and similarly two $\Upsilon$ type states. The chirality projection again relates
$\Psi$ and $\Upsilon$ type spinors, so that we are left with $2$ SUSY from this sector
labeled by $s_1$. Collecting all, one obtains $6$ or 3d $\mathcal{N}=3$ SUSY.

Considering the D1-brane projection, again we take $A=\pm 1$ and $F=\pm 1$ components.
From the first $4$ SUSY of the $\mathcal{N}=3$ theory, one obtains
$(s,s;,s,-s;s,s)$ or $(s,-s;-s,-s;-s,s)$ with $s=\pm$, obtaining
$\Psi_{s,s,s}+\Upsilon_{-s,s,s}$ after the chirality projection. However, in
the last set of $2$ SUSY of the $\mathcal{N}=3$ theory, note that different
$F$ eigenstates are all mixed up for given value of $A$ eigenvalue.
As this makes it impossible to correlate the signs of $A$ and $F$ eigenvalues,
D1-branes cannot preserve this part of SUSY. So we have D1-brane vortices preserving
$2$ SUSY. SUSY of fundamental string vortices can be studied similarly.
Its projection $\sigma_3\otimes\Gamma^{05}=AD$ demands $A=-D=\pm 1$, where the relative
minus sign is chosen to stay in the same BPS sector as D1-branes. From $4$ of $\mathcal{N}=3$
SUSY, one obtains $\Psi_{s,s,s}+\Upsilon_{-s,s,s}$, which are the same $2$ SUSY as those for
D1-branes. From $2$ of the $\mathcal{N}=3$ SUSY, again no further SUSY appears.

From field theory, the supersymmetry of the $\mathcal{N}=3$ theory is obtained
by restricting the $\mathcal{N}=4$ SUSY by identifying $SU(2)_L$, $SU(2)_R$, and
taking the same off-shell (for three D-term fields) SUSY for $Q^{ab}_\alpha$
for symmetric $a,b$. Equivalently, one can write the supercharges as $Q^A_\alpha$.
The $2$ SUSY preserved by our vortices take the form of $Q^{11}_-\sim(Q^{22}_+)^\dag$,
and this will be the same $2$ supercharges that we will use to calculate the index even
in the $\mathcal{N}=4$ theories. The BPS equations for the $\mathcal{N}=3$
topological vortices are the same as those for the $\mathcal{N}=4$ vortices.

The fundamental strings discussed above should be distinguished from the topological 
vortices in \textit{classical} field theory, as the so-called non-topological vortices.
However, only the total spectrum of all vortices will have a duality invariant meaning 
in partially unbroken phases. Non-topological vortices are discussed in the literatures: 
for instance, see \cite{Eto:2010mu} and references therein. In particular, non-topological
vortices in supersymmetric Maxwell-Chern-Simons theories are studied in \cite{Jackiw:1990pr}. 
The mass is given by the electric charge in the unbroken phase multiplied by the mass 
of an elementary particle \cite{Jackiw:1990pr}, supporting that
they are bounds of fundamental strings.

\subsection{Vortex quantum mechanics}

We review the quantum mechanical description of topological BPS vortices in the
$\mathcal{N}\!=\!4$ theory \cite{Hanany:2003hp} motivated by branes, and also explain
how to include the effect of nonzero Chern-Simons term preserving $\mathcal{N}\!=\!3$
supersymmetry \cite{arXiv:0805.0602,Collie:2008za}. As explained before, it has been discussed
\cite{Shifman:2006kd,Eto:2006uw} that some of the degrees in this mechanics come from
non-normalizable zero modes of the soliton, demanding special care about IR regularization
to correctly understand their low energy dynamics \cite{Shifman:2006kd}. As concretely
supported with single vortices and generally argued in the previous subsection,
we think the difference between the two mechanical models will not affect the index that
we calculate and study, by having two models connected by a continuous supersymmetric
deformation (zooming into region of the moduli space with minimal sizes).

In the $\mathcal{N}\!=\!4$ theory, the $4$ supercharges $Q^{a\dot{1}}_-$ (and the conjugate
$Q^{a\dot{2}}_+$) preserved by the vortices appear as the supercharges of the mechanical
model. We call $Q^a\equiv Q^{a\dot{1}}_-$ in the mechanics. The $SU(2)_L$ global
symmetry (with $a$ doublet index) is manifest. As explained in \cite{Hanany:2003hp},
the dynamical degrees of this mechanics can be obtained
by a dimensional reduction of 4d $\mathcal{N}=1$ superfields down to 1d, regarding
the above $SU(2)_L$ as the internal 3d rotation in the 4d to 1d reduction. $SU(2)_R$ in the
3d QFT is broken by the FI term to $U(1)_R$. As the hypermultiplet scalar $q^i\equiv q^{i\dot{1}}$
assumes nonzero expectation value, the surviving $U(1)$ is a linear combination of $U(1)_R$
and the overall $U(1)$ of $U(N)$ gauge symmetry which leave the VEV invariant. We simply
call the last combination $U(1)_R$.

The gauged quantum mechanics for $k$ vortices has the following degrees:
$N$ chiral multiplets $q^i,\psi^{ia}$ in the fundamental representation of $U(k)$,
$N_f-N$ chiral multiplets $\tilde{q}_p,\psi^a_p$ in the anti-fundamental representation
of $U(k)$, a chiral multiplet $Z,\chi_a$ in the adjoint representation of $U(k)$,
and the $U(k)$ vector multiplet $A_t,\phi^I,\lambda_a$. The variables $q^i$ and $\tilde{q}_p$
should not be confused with complex scalar fields in 3d QFT. In fact, the moduli coming from
these mechanical variables all originate from the zero modes of $q$ fields in QFT.
The Lagrangian is given by \cite{Hanany:2003hp}
\begin{eqnarray}\label{QM-action}
  L&=&{\rm tr}\left[\frac{1}{2}D_t\phi_ID_t\phi_I+|D_tZ|^2+|D_tq|^2+|D_t\tilde{q}|^2 +i\bar\lambda^a D_t\lambda_a +i\bar\chi^a D_t\chi_a+i \bar\psi^a D_t\psi_a+i \bar{\tilde\psi}^a D_t\tilde\psi_a
  \right.\nonumber\\
  &&+\frac{1}{4}[\phi_I,\phi_J]^2-|[\phi_I,Z]|^2-qq^{\dagger}\phi_I\phi_I - \tilde{q}^{\dagger}\tilde{q}\phi_I\phi_I
  -\frac{1}{2}\big([Z,Z^\dagger]+qq^\dagger-\tilde{q}^\dagger \tilde{q}-r\big)^2  \nonumber \\
  &&+\bar\lambda^a(\sigma^I)_{ab}[\phi_I,\lambda^b]+\bar\chi^a(\sigma^I)_{ab}[\phi_I,\chi^b] +\bar\psi^a(\sigma^I)_{ab}\phi_I\psi^b-\bar{\tilde\psi}^{a}(\sigma^I)_{ab}\tilde{\psi}^b\phi_I
  \nonumber \\
  &&\left.+\sqrt{2}i\left(\bar\chi^a[\bar\lambda_a,Z]+[Z^\dagger,\lambda^a]\chi_a+q\bar\psi^a
  \bar\lambda_a+\lambda^a\psi_a q^\dagger-\tilde{q}^\dagger\tilde\psi^a\lambda_a
  - \bar{\lambda}^a\bar{\tilde\psi}_a\tilde{q}\right)\right]\ ,
\end{eqnarray}
where all $SU(2)_L$ doublet indices are raised/lowered by $\epsilon^{ab},\epsilon_{ab}$.
The $N$ chiral multiplet fields are regarded as $k\times N$ matrices, while $N_f-N$ of them
with tilde are regarded as $(N_f-N)\times k$ matrices. $r$ is proportional to the inverse of
3d coupling constant, $\frac{1}{g_{YM}^2}$. The supersymmetry and other properties of this
model is summarized in Appendix A. The classical solution for the ground state is given by
taking the D-term potential to vanish,
\begin{equation}
  [Z,Z^\dag]+q^iq^\dag_i-\tilde{q}^{p\dag}\tilde{q}_p=r\ .
\end{equation}
In the D-brane realization, the sign of $r$ depends on the relative position
of the two NS5-branes in Fig \ref{n=4-brane}. On the left side of the figure, the vortex mechanics
for the corresponding 3d theory has $r\!>\!0$. On the right side, $r\!<\!0$ for the putative
Seiberg-dual theory. The moduli spaces of the vortices are different for $r\gtrless 0$, but
their real dimensions are all $2N_f k$.

The effect of nonzero Chern-Simons term to this mechanics is investigated in
\cite{Kim:2002qma}, and more recently in \cite{arXiv:0805.0602,Collie:2008za}.
To the above gauged quantum mechanics, we add the following term \cite{Collie:2008za}
\begin{equation}\label{CS-mechanics}
  \Delta L=\kappa\ {\rm tr}(A_t+\phi)
\end{equation}
where $\phi\equiv\phi^3$ is the component of the vector multiplet scalar
along the nonzero FI parameter $\zeta=\zeta^3$. (\ref{CS-mechanics}) is argued to encode
the correction in the moduli space dynamics to the leading order in $\kappa$
\cite{arXiv:0805.0602,Collie:2008za}. So this model should be reliable (of course modulo
the non-normalizable mode effects) when the Yang-Mills mass scale $\kappa g_{YM}^2$ is
much smaller than the FI mass scale $\zeta$. Again, the Witten index we study in this
paper does not depend on such continuous
parameters, which justifies our usage of this model for calculating the index.

The term (\ref{CS-mechanics})
breaks $4$ SUSY of the $\mathcal{N}=4$ vortices to $2$, as it should for our
$\mathcal{N}=3$ vortices. To see this, recall the supersymmetry
transformation of appendix A,
\begin{equation}
    Q_a A_t= i \bar\lambda_a \ , \quad \bar{Q}_a A_t = -i\lambda_a \ ,\ \
    Q_a \phi^I = i(\tau^I)_a^{\ b}\bar\lambda_b \ , \quad \bar{Q}_a\phi^I =
    i (\tau^I)_a^{\ b}\lambda_b \ . \nonumber
\end{equation}
The term (\ref{CS-mechanics}) only preserves $Q_2\sim Q^{1}$ and complex conjugate
$\bar{Q}_1$, since $(\tau^3)_1^{\ 1}=-(\tau^3)_2^{\ 2}=1$. (\ref{CS-mechanics})
also breaks $SU(2)_L$ to $U(1)$, which should happen as the two $SU(2)$ R-symmetries are
locked in the $\mathcal{N}\!=\!3$ theories, broken to $U(1)_R$ by the FI term.

Perhaps it is also worthwhile to emphasize that this model was originally considered
in \cite{arXiv:0805.0602,Collie:2008za} as vortex quantum mechanics of $\mathcal{N}=2$
theories. At the level of classical field theory, the difference of the $\mathcal{N}=2$
theory considered there and our $\mathcal{N}=3$ theory is that the latter has an extra
term coming from a nonzero superpotential which couples $\phi^1+i\phi^2$ to $q\tilde{q}$.
Any possible difference in the vortex moduli space dynamics coming from this superpotential
should appear always with the 3d field $\tilde{q}_i$, which are always set to zero for
classical vortex solutions. Thus, the bosonic part of the quantum mechanics (consisiting of
the vortex zero modes) will never be affected. The only possible issue is the fermionic term
proportional to $\kappa$, which may be added in the case of $\mathcal{N}=3$ vortices,
separately preserving the same $2$ SUSY. This will be a well-defined problem which can be
studied with the SUSY transformation of appendix A.

Although we have not carefully studied this possibility, the overall coefficient of
the extra fermionic term is not constrained by the $2$ SUSY of vortex quantum mechanics only.
So we should be able to deform the mechanics model in a continuous way preserving supersymmetry,
turning off this term. Then, the possible difference will not affect the index we study.
However, there could possibly be an important
difference between the two models,
as the two quantum mechanics models for $\mathcal{N}=2$ and $3$ theories may come with
different values of $U(1)_R$ charges. This ambiguity appears because the mechanics only
has a D-term potential without an F-term potential. The value of this charge $R$ for the
$U(k)$ fundamental variables $q^i,\tilde{q}_p$ are left undetermined in the index
calculation. For $\mathcal{N}=3$ vortices, we should plug in the canonical value
$R=\frac{1}{2}$ inherited by the zero modes of 3d fields. For $\mathcal{N}=2$ vortices,
there is a possible anomalous shift of $R$ in 3d matter fields, which is meaningful at least
at the conformal point with $\zeta\!=\!0$. If one studies $\mathcal{N}\!=\!2$ vortices to
probe the physics at the conformal point, it may be important to take the R-charge as that
of the IR CFT with $\zeta=0$. In this paper, we only consider the $\mathcal{N}=3$ version
of the index.

\subsection{$\mathcal{N}=4$ and $3$ indices for vortices}

\begin{table}[t!]
$$
\begin{array}{c|cccc|cc|c}
  \hline &q&\psi_a&\tilde{q}&\tilde\psi_a&Z&\chi_a&Q_a\\
  \hline SO(2,1)&0&-1/2&0&-1/2&1&1/2&-1/2\\
  U(1)_R&R&R+1/2&\tilde{R}&\tilde{R}+1/2&0&1/2&1/2\\
  SU(2)_L&0&\pm 1/2&0&\pm 1/2&0&\pm 1/2&\pm 1/2\\
  \hline U(N)&\bar{N}&\bar{N}&1&1&1&1&1\\
 U(N_f-N)&1&1&N&N&1&1&1\\
  \hline
\end{array}
$$
\caption{Global charges of mechanical variables for $\mathcal{N}=4$ vortices}\label{charge}
\end{table}
To define and study a Witten index partition function for topological vortices,
we discuss the symmetries of the vortex quantum mechanics in more detail.
Consider the $\mathcal{N}=4$ vortex first. The $SU(2)_L$ of the mechanics is inherited
from the 3d QFT. We denote its Cartan by $J_L$, whose values for
mechanical variables are given in Table \ref{charge}. Our convention is that
the upper $a=1$ component has $J_L=+\frac{1}{2}$, and so on. There is also an $SO(2)$
symmetry which rotates $Z$ with charge $1$. As the diagonals of $Z$ roughly correspond to
$k$ positions of vortices, we consider it as the rotational symmetry of the 3d theory
in $SO(2,1)$. We call this charge $J_E$, whose values are listed on the first row of
Table \ref{charge}. The charges for $q,\tilde{q}$ are taken to be
zero because they come from the internal zero modes. Once the charges of
bosonic variables are determined, their superpartners' charges are fixed by noting that
$Q^a$ comes from $Q^{a\dot{1}}_-$ of 3d QFT, which has $J_E=-1/2$.
Finally, we consider $U(1)_R$ charge $J_R$ which is inherited from the
unbroken Cartan of $SU(2)_R$. We want $Z$ to be neutral. $q$, $\tilde{q}^\dag$
subject to the mechanical D-term constraint form the internal moduli space of vortices.
In 3d solitons, they appear partly from the $U(1)$ embedding of the ANO vortex
into $U(N)$ (for $q$'s), and also because asymptotic VEV for hypermultiplet fields can be
different from the value at the core of each vortex (for $\tilde{q}^\dag$). So in QFT,
these moduli all come from the $N\times N_f$ fundamental hypermultiplets (which we also
called $q$ in 3d), by decomposing them into $N\times N$ and $N\times (N_f-N)$.
From the unbroken global symmetry, it seems clear that $k\times N$ scalar $q^i$ and
$k\times (N_f-N)$ scalar $\tilde{q}^{p\dag}$ in mechanics should have same $J_R$ charge.
So in Table \ref{charge}, we naturally set $\tilde{R}=-R$. We shall mostly keep $R$,
$\tilde{R}$ as unfixed parameters in general considerations, but at various final stages
set $R=-\tilde{R}$.\footnote{It happens that
the remaining value $R$ will never appear in the $\mathcal{N}\!=\!4$ index, assuming
$\tilde{R}=-R$. For $\mathcal{N}\!=\!3$ vortices, the index will depend on $R$ even after
setting $\tilde{R}=-R$.} Furthermore, from
the fact that this $U(1)_R$ is inherited from 3d $U(1)_R\subset SU(2)_R$, we expect
$R=\frac{1}{2}$ for $\mathcal{N}=4,3$ theories.

Now we consider the Witten index
\begin{equation}\label{index-trace}
  I_k(\mu_i,\gamma,\gamma^\prime)={\rm Tr}_k\left[(-1)^Fe^{-\beta Q^2}e^{-\mu^i\Pi_i}
  e^{-2i\gamma J}e^{-2i\gamma^\prime J^\prime}\right]
\end{equation}
for $\mathcal{N}=4$ vortices, where $J\equiv J_R+J_L+2J_E$,
$J^\prime=J_R-J_L$. This index counts states preserving $Q^1$ in mechanics, or
$Q^{1\dot{1}}_-$ in QFT. $J_R+J_L$ appearing in $J$ is an
$\mathcal{N}=2$ R-charge, which is the first $12$ plane rotation in $SO(4)$. $J_R-J_L$
chemical potential $\gamma^\prime$ has to be turned off when we try to understand the
$\mathcal{N}=2$ SUSY structure of the index, and also for $\mathcal{N}\!=\!3$ vortices.
The trace is taken over the Hilbert space of all single- or multi-particle states
with vorticity $k$. $\beta$ is the usual regulator in the Witten index and does
not appear in $I_k$. Finally, $\Pi_i$ for $i=1,2,\cdots,N_f$ are the $S[U(N)\times U(N_f-N)]$
Cartan charges, subject to the condition that $\sum_i\Pi_i$ is a gauge symmetry.
In the mechanical model, this overall $U(1)$ is absorbed into the overall $U(1)$ of
$U(k)$ gauge symmetry.

Considering the Euclidean path integral expression for the above index,
the chemical potentials $\gamma$, $\mu^i$ provide regulating mass terms for the
zero modes of the vortex mechanics. $\gamma$ is well known as the Omega deformation
of the spatial rotation. The index interpretation of the $\gamma$ dependent part
is well understood. In particular, the degree of divergence of each term of the index
as one takes $\gamma\rightarrow 0$ is naturally interpreted as the particle number
of the states. See a detailed explanation of \cite{Kim:2011mv} in the context of
5d instanton bound state counting, which applies to our case as well.
The $\mu_i$ dependent part of the index however seems subtle and needs a proper
interpretation, as they correspond to internal zero modes. We do not have a good
physical interpretation at the moment. See the later part of this subsection for
a more detailed explanation on why it is subtle for $N_f>N$.

This index can be calculated by using localization technique \cite{Shadchin:2006yz},
similar to that used to calculate the instanton partition functions in 4d
or 5d gauge theories. In appendix A, we explain a slightly unconventional
calculation, which is perhaps a bit more straightforward in that there is no need
for a contour prescription appearing in `standard' calculations. Of course we shall
also view our result in the standard context, using contour integrals. The localization
calculation consists of identifying the saddle points, and then calculating the
determinants around them. In appendices B and C, we illustrate the calculation for
$k\!=\!1$. We also checked the formulae below for some higher $k$'s.

The saddle points for the $k$ vortex index in our calculation are labeled by the
so-called one dimensional $N$-colored Young diagrams with box number $k$. It is
obtained by dividing $k$ into $N$ different non-negative integers
\begin{equation}\label{young}
  k=k_1+k_2+\cdots+k_N\ ,
\end{equation}
where $N$ non-negative integers $k_i$ are ordered. The index contribution from
the saddle point $(k_1,k_2,\cdots, k_N)$ is given by
\begin{eqnarray}\label{general-index}
   I_{(k_1,k_2,\cdots,k_N)}=\prod_{i=1}^{N}\prod_{s=1}^{k_i}\left[\prod_{j=1}^{N} \frac{\sinh\frac{E_{ij}-2i(\gamma-\gamma^\prime)}{2}}{\sinh\frac{E_{ij}}{2}}
   \prod_{p=N+1}^{N_f}\frac{\sinh\frac{E'_{ip}-2i(\gamma+\gamma^\prime)(R+\tilde{R})
   +2i(\gamma-\gamma^\prime)}{2}}{\sinh\frac{E'_{ip}-2i(\gamma+\gamma^\prime)
   (R+\tilde{R})}{2}}\right]
\end{eqnarray}
where
\begin{eqnarray}
    E_{ij} = \mu_i-\mu_j+4i\gamma(k_j-s+1)\,,
    \quad E'_{ij} = \mu_i-\mu_j-4i\gamma(s-1)\ .
\end{eqnarray}
This expression also admits a contour integral expression:
\begin{eqnarray}\label{contour}
  I_k&=&\frac{1}{(2i)^kk!}\oint\prod_{I=1}^k\left[\frac{d\phi_I}{2\pi}\prod_{i=1}^N
  \frac{\sinh\frac{\phi_I-\mu_i+2i(\gamma+\gamma^\prime)R-2i(\gamma-\gamma^\prime)}{2}}
  {\sinh\frac{\phi_I-\mu_i+2i(\gamma+\gamma^\prime)R}{2}}\prod_{p=N+1}^{N_f}
  \frac{\sinh\frac{\phi_I-\mu_p-2i(\gamma+\gamma^\prime)\tilde{R}+2i(\gamma-\gamma^\prime)}{2}}
  {\sinh\frac{\phi_I-\mu_p-2i(\gamma+\gamma^\prime)\tilde{R}}{2}}\right]\nonumber\\
  &&\times\prod_{I\neq J}\sinh\frac{\phi_{IJ}}{2}\prod_{I,J}
  \frac{\sinh\frac{\phi_{IJ}+2i(\gamma+\gamma^\prime)}{2}}{\sinh\frac{\phi_{IJ}+4i\gamma}{2}
  \sinh\frac{\phi_{IJ}-2i(\gamma-\gamma^\prime)}{2}}\ .
\end{eqnarray}
The integration contour has to be carefully chosen so that only a subset of residues
in the integrand are kept. Introducing $z_I=e^{\phi_I}$, the contour for
$z_I$ takes the form of a closed circle. The simplest possible choice might have been a
unit circle surrounding the origin $z_I=0$, regarding $\phi_I$ as $i$ times a $2\pi$
periodic angle. The contour is actually more complicated than this. It has to be
chosen in a way that the poles coming from the $\prod_{p=N+1}^{N_f}$ product of
(\ref{contour}) all stay outside the contour circle. Also, the poles from
$\sinh\frac{\phi_{IJ}-2i(\gamma-\gamma^\prime)}{2}$ on the second line as well as
poles at $z_I=0$ coming from $d\phi_I=\frac{dz_I}{z_I}$ are
taken outside the contour. Such an exclusion of some residues is also familiar in the
instanton calculus with complicated matter contents. We explicitly checked this
statement on the contour for some low values of $k$ and $N,N_f$.

By carefully considering the above contour integration expression, one can decompose
this index to various contributions from different $\mathcal{N}\!=\!2$ supermultiplets.
Firstly, $\mathcal{N}\!=\!4$ vector multiplet combines with the $N\times N$ part of the
$N\times N_f$ hypermultiplets (which assume nonzero asymptotic VEV)  to yield a
basic contribution. In the $\mathcal{N}=2$ language, these contributions can
be decomposed into those from one vector supermultiplet, one adjoint chiral multiplet
(participating in the $\mathcal{N}=4$ vector multiplet), and $N^2$ extra fundamental
chiral multiplets and anti-fundamental chiral multiplets. The contributions are given by
\begin{equation}
  \hspace*{-0.3cm}
  z_{v}=\prod_{j=1}^N\frac{\sinh\frac{E_{ij}^\prime}{2}}{\sinh\frac{E_{ij}}{2}},\
  z_{adj}=\prod_{j=1}^N\frac{\sinh\frac{E_{ij}-2i(\gamma-\gamma^\prime)}{2}}
  {\sinh\frac{E_{ij}^\prime-2i(\gamma-\gamma^\prime)}{2}},\
  z^{N}_{fund}=\prod_{j=1}^N\frac{1}{\sinh\frac{E_{ij}^\prime}{2}},\
  z^{N}_{anti}=\prod_{j=1}^N\sinh\frac{E_{ij}^\prime-2i(\gamma-\gamma^\prime)}{2}\ .
\end{equation}
The index contribution from this sector is the product of all these four factors.
The `antichiral' part denotes contribution from the $N\times N$ block of the
anti-fundamental superfields $\tilde{q}_i$, which contribute only to the fermion zero
modes without bosonic zero modes.\footnote{The fermions are superpartners of
bosonic zero modes only for vortices preserving $4$ SUSY in $\mathcal{N}\!=\!4$ theories.}
Consider the combinations of the two contributions $z_vz^N_{fund}$ and $z_{adj}z^N_{anti}$:
\begin{equation}
  z_{v}z^N_{fund}=\prod_{j=1}^N\frac{1}{\sinh\frac{E_{ij}}{2}}\ ,\ \
  z_{adj}z^N_{anti}=\prod_{j=1}^N\sinh\frac{E_{ij}-2i(\gamma-\gamma^\prime)}{2}\ .
\end{equation}
The first part $z_vz^N_{fund}$ is called $z^{\rm vortex}$ in
\cite{Dimofte:2010tz} for 2d $U(1)$ theories (i.e. $N=1$). In this case, the term $\mu_i-\mu_j$
in $E_{ij}$ is simply ignored and $\gamma^\prime=0$ as we ignore $SU(2)_L$.
Also, as there is only one 1d Young diagram of length $k$ in the $U(1)$ case, the product of
$s$ simply runs over $s=1,2,\cdots,k$. Rescaling all chemical potentials as
\begin{equation}
   ({\rm 3d\ chemical\ potentials})=\beta({\rm 2d\ parameters})
\end{equation}
and taking $\beta\rightarrow 0$ as the 2d limit, keeping all 2d parameters fixed,
one obtains
\begin{equation}
  z^{\rm vortex}=\prod_{s=1}^k\frac{1}{2i\gamma(k-s+1)}=\frac{1}{k!\hbar^k}\ ,
\end{equation}
where $\hbar\equiv 2i\gamma$, apart from $\beta$ dependent factor which in our case
cancels with other contributions (and in $\mathcal{N}=2$ theories like \cite{Dimofte:2010tz}
should be absorbed into the fugacity $q$ for vorticity). This agrees with \cite{Dimofte:2010tz}.
The extra part $z_{adj}z^N_{anti}$ seems to be unexplored in the $\mathcal{N}=2$ context.

For $N_f\!>\!N$, we also have extra contributions from $N_f-N$ hypermultiplets,
which decomposes to $N_f-N$ fundamental and anti-fundamental chiral multiplets.
From our $\mathcal{N}=4$ formula, the contributions of these two are
\begin{equation}
  z^{N_f-N}_{fund}=\prod_{p=N+1}^{N_f}\frac{1}{\sinh\frac{E_{ip}^\prime}{2}}
  \ ,\ \ z^{N_f-N}_{anti}=\prod_{p=N+1}^{N_f}\sinh\frac{E_{ip}^\prime
  +2i(\gamma-\gamma^\prime)}{2}\ ,
\end{equation}
where the `anti-chiral' contribution again denotes that from the
fermion zero modes of $\tilde{q}_p$ for $p=N+1,\cdots, N_f$.
As explained above, we took $R=-\tilde{R}$.
To compare these with 2d $\mathcal{N}=2$ results, we take $\gamma^\prime=0$.
Let us identify the Scherk-Schwarz masses of the fields $q^i$, $\tilde{q}_i^\dag$ from
the chemical potentials, as we reduce the theory to 2d. The masses are proportional to
\begin{eqnarray}
  q^i&:&\mu_i(1)+\mu_p(-1)+2i\gamma R+2i\gamma^\prime R=
  \mu_{ip}+2iR(\gamma+\gamma^\prime)\nonumber\\
  (\tilde{q}_i)^\dag&:&\mu_i(1)+\mu_p(-1)+2i\gamma(-R)+2i\gamma^\prime(-R)=
  \mu_{ip}-2iR(\gamma+\gamma^\prime)\ .
\end{eqnarray}
Taking $-\frac{1}{2}$ times these to be the masses,
the fundamental and the anti-fundamental chiral multiplets have
the following difference in their masses:
\begin{equation}
  m_{\tilde{q}}=m_{q}+2iR(\gamma+\gamma^\prime)\rightarrow m_q+i(\gamma+\gamma^\prime)\ .
\end{equation}
In particular, in the $U(1)$ case ($N=1$), one finds (with $\gamma^\prime=0$)
\begin{eqnarray}
  z_{fund}^{N_f-N}&=&\prod_{s=1}^k\frac{1}{\frac{\mu_{ip}}{2}-2i\gamma(s-1)}=
  \prod_{s=1}^k\frac{1}{-m_q-2i\gamma(s-1)-i\gamma/2}\label{chiral}\\
  z_{anti}^{N_f-N}&=&\prod_{s=1}^k(\frac{\mu_{ip}}{2}-2i\gamma(s-1)+i\gamma)=
  \prod_{s=1}^k(-m_{\tilde{q}}-2i\gamma(s-1)+3i\gamma/2)\label{anti-chiral}\ .
\end{eqnarray}
The analogous $\mathcal{N}\!=\!2$ result of \cite{Dimofte:2010tz} is\footnote{We corrected
the ranges of $s$ summation in eqns.(2.29) and (2.30) of \cite{Dimofte:2010tz},
based on their eqns.(2,23), (2.27) and (2.28). In any case, this difference can be
absorbed by an overall shift of all masses by $\hbar$.}
\begin{equation}\label{matter-DGH}
  z_{fund}=\prod_{s=1}^k\frac{1}{m+(s-1)\hbar}\ ,\ \ z_{anti}=
  \prod_{s=0}^{k-1}(m+(s-1)\hbar)
\end{equation}
for a given mass $m$ for a chiral or anti-chiral mode. Up to an overall shift
of our masses by $-i\gamma/2$, this is same as our result, up to factors of $-1$
which in our case all cancel out in (\ref{general-index}).

It would be illustrative to take a more detailed look at the
formula for single vortices, to explain the index interpretation in some cases
and also to emphasize a subtlety. From (\ref{young}), one has $N$ different
saddle points. The total index at single vorticity is thus given by their sum:
\begin{equation}\label{single-index-n4}
  I_{k=1}=\frac{\sin(\gamma+\gamma')}{\sin2\gamma}
    \sum_{i=1}^N\prod_{j(\neq i)}^{N}\frac{\sinh\left(\frac{\mu_{ji}+2i(\gamma-\gamma')}{2}\right)}
    {\sinh\frac{\mu_{ji}}{2}}
    \!\!\prod_{p=N+1}^{N_f}\!\!\frac{\sinh\left(\frac{\mu_{pi}-2i(\gamma-\gamma')}{2}\right)}
    {\sinh\frac{\mu_{pi}}{2}}\ .
\end{equation}
From the calculation of 1-loop determinants at $k=1$ in appendix C,
one can easily show that the factor $\frac{\sin(\gamma+\gamma^\prime)}{\sin 2\gamma}$
combines the contribution $\sin^{-2}(2\gamma)$ from the center-of-mass zero
modes and $\sin 2\gamma\sin(\gamma+\gamma^\prime)$ from the Goldstone fermion zero modes
for $4$ broken supercharges. $\gamma,\gamma^\prime$ are lifting, or regularizing, these
zero modes. So we shall call them as the center-of-mass index
\begin{equation}\label{com}
  I_{\rm com}=\frac{\sin(\gamma+\gamma^\prime)}{\sin 2\gamma}
\end{equation}
for a single super-particle. To get the real information on bound state degeneracies, one
has to expand the denominator in certain powers of the fugacity $e^{i\gamma}$ and extract out
their (integral) coefficients. Just like the instanton index in 5d theories studied
in \cite{Kim:2011mv}, expanding (\ref{com}) in the fugacity is ambiguous. However, just
as in \cite{Kim:2011mv}, it suffices to identify a factor of (\ref{com}) as accompanying
the translation degree per super-particle, factored out from the more informative internal
degeneracy factor. In fact, as we shall explain in
more detail in the next section, one always obtains a single factor of (\ref{com})
when one extracts out the single particle partition function from the general
multi-particle result (\ref{general-index}). This factor is also ignored in all
sorts of bound state counting with translational zero modes.

The remaining factor in (\ref{single-index-n4}) is more nontrivial. For the case
with $N_f=N$, namely for local vortices, this remainder becomes very simple after
one sums over $N$ saddle points. With explicitly summing over them for a few
low values of $N$, one can easily confirm that
\begin{equation}
  I_{k=1}=I_{\rm com}(\gamma,\gamma^\prime)\left(e^{i(N-1)(\gamma-\gamma^\prime)}
  +e^{i(N-3)(\gamma-\gamma^\prime)}+\cdots+e^{-i(N-1)(\gamma-\gamma^\prime)}\right)
  \equiv I_{\rm com}\ \chi_{N}(\gamma-\gamma^\prime)
\end{equation}
for $N_f\!=\!N$. The number of states $\chi_{N}(0)=N$ from the internal
part of moduli space is finite, which is simply due to the compactness of
the internal moduli space $\mathbb{CP}^{N-1}$ for local vortices. So, as the trace
expression (\ref{index-trace}) obviously implies, the index for $N_f\!=\!N$ can be
naturally regarded as a Witten index counting degeneracy.

For semi-local vortices with $N_f\!>\!N$, the remainder of (\ref{single-index-n4}) is
subtler. Unlike the case with $N_f\!=\!N$, the flavor chemical potentials $\mu_i$
survive even after one sums over $N$ saddle points. In particular, the dependence
on $\sinh\frac{\mu_{ij}}{2}$ in the denominator survives in the index,
making its expansion in the fugacities $e^{\mu_i}$ again ambiguous like (\ref{com}).
Just like (\ref{com}), the chemical potentials $\mu_i$ regularizes the internal zero
modes, some of them being noncompact for $N_f\!>\!N$. For the index for 5d instanton particles,
two interpretations were provided to such an internal index in different contexts \cite{Kim:2011mv}.
Firstly in the Coulomb branch in which $U(N)$ is broken to $U(1)^N$, the signs of
the Noether charges for $U(1)^N$ are fixed in a BPS sector, making the expansion of the
denominator unambiguous. Secondly, in the symmetric phase, in which the
whole $U(N)$ gauge symmetry is unbroken, the same index was proven to be a
superconformal index which counts gauge invariant operators, after a well-defined $U(N)$
singlet projection. Here for semi-local vortices, it seems that the vortex partition
function is similar to neither of the two cases. As $S[U(N)\times U(N_f-N)]$ is a global
symmetry, the gauge invariance projection is unnecessary. Also, since this symmetry is
unbroken, there is no fixed sign for their Cartans either. Rather, one should expand the
expression (\ref{single-index-n4}) into the irreducible characters of the global symmetry.
We have attempted this expansion of (\ref{single-index-n4}). It does not clearly work in
an unambiguous way, essentially due to an ambiguity on how to go around the poles in
(\ref{single-index-n4}). So for semi-local vortices, we do not have a clear understanding
of its Witten index interpretation, despite its formal expression (\ref{index-trace}) as
trace over Hilbert space. Perhaps a new interpretation of noncompact internal modes
from a parton-like picture \cite{Collie:2009iz} might be necessary.

Even without a solid index interpretation, we can get useful information from them,
regarded a kind of supersymmetric partition functions on $\mathbb{R}^2\times S^1$.
In the next section, we use them to study Seiberg dualities. The index interpretation
helps when available, but is not essential.

The index for the $\mathcal{N}\!=\!3$ theory turns out to be very similar to the above
$\mathcal{N}=4$ index, with small changes. From the quantum mechanics analysis,
we obtain the general formula
\begin{eqnarray}\label{general-cs-index}
   I_{\{k_1,k_2,\cdots,k_N\}}=e^{-S_0}\prod_{i=1}^{N}\prod_{s=1}^{k_i}\left[\prod_{j=1}^{N} \frac{\sinh\frac{E_{ij}-2i\gamma}{2}}{\sinh\frac{E_{ij}}{2}}\prod_{p=N+1}^{N_f}
    \frac{\sinh\frac{E'_{ip}-2i\gamma(R+\tilde{R})+2i\gamma}{2}}
    {\sinh\frac{E'_{ip}-2i\gamma(R+\tilde{R})}{2}}\right]
\end{eqnarray}
with same definitions of $E_{ij}$, $E_{ip}^\prime$, and
\begin{eqnarray}
e^{-S_0} = e^{-\kappa\sum_{i=1}^N\sum_{s=1}^{k_i}[\mu_i-2i\gamma R-4i\gamma(s-1)]}\ .
\end{eqnarray}
Note that the $U(1)_R$ charge $R$ does appear in $S_0$  part of the index even after setting
$\tilde{R}=-R$. We set $R=1/2$ for the $\mathcal{N}=3$ theory.

\section{Seiberg dualities}

\subsection{$\mathcal{N}=4$ dualities from vortices}

Seiberg dualities for $\mathcal{N}\!=\!4$ gauge theories in a naive form are
motivated by branes \cite{Kapustin:2010mh}. Consider the brane configuration
on the left side of Fig \ref{n=4-brane} with $\zeta=0$, where NS5$^\prime$-brane
is not displaced relative to the NS5-brane in $x^7,x^8,x^9$ directions.
The resulting $U(N)$ $\mathcal{N}=4$ gauge theory is coupled to $N_f$ ($\geq N$)
fundamental hypermultiplets. At low energy, this theory may (but not always) flow to a
superconformal field theory. Now consider the configuration obtained by letting
the two NS5-branes to cross each other by moving along $x^6$ direction. By the brane
creation effect \cite{Hanany:1996ie}, on the right side of the figure there are $N_f-N$
D3-branes stretched between two NS5-branes. So one obtains an $\mathcal{N}=4$ $U(N_f-N)$
gauge theory coupled to $N_f$ matters. Supposing that both theories flow to SCFT,
one would have obtained a brane realization of two QFT with same IR fixed point, and thus
a Seiberg-like duality. However, as shown in \cite{Gaiotto:2008ak}, this happens only under
a restrictive condition.
The main method of \cite{Gaiotto:2008ak} is to study the R-charges of BPS magnetic
monopole operators in the UV theory, and see if they can sensibly saturate the
superconformal BPS bounds for the scale dimensions of local operators in IR. Picking
an $\mathcal{N}=2$ R-charge $R$ (given by $J_R+J_L$ in the notation of our previous
section), the BPS scale dimensions $\Delta$ of chiral monopole operators saturate
the bound
\begin{equation}
  \Delta\geq R\ .
\end{equation}
From the unitarity bound, the scale dimensions these operators
should satisfy $\Delta\geq \frac{1}{2}$. So if any of the monopole operators have
$R$ smaller than $\frac{1}{2}$, the QFT cannot flow to an $\mathcal{N}=4$ SCFT, at least
not in a way that uses the UV $SO(4)$ R-charges as the superconformal R-charges in IR.
\cite{Gaiotto:2008ak} refers to this as the absence of `standard IR fixed point.'

Considering the monopole operator with the $U(N)$ GNO charge $H=(n_1,n_2,\cdots,n_N)$,
with integer entries, one obtains the following R-charge
\begin{equation}\label{n=4-R-charge}
  R=\frac{N_f}{2}\sum_{i=1}^N|n_i|-\sum_{i<j}|n_i-n_j|
\end{equation}
of the monopole operator. Plugging in $H=(1,0,0,\cdots,0)$ charge, one obtains a simple
necessary condition
\begin{equation}\label{n=4-unitarity}
  \frac{N_f}{2}-N+1\geq\frac{1}{2}\ \rightarrow\ \ N_f\geq 2N-1
\end{equation}
for the existence of a standard fixed point. Indeed, if this condition is satisfied,
there are no violations of the unitarity bound for other monopole operators
\cite{Gaiotto:2008ak}.

Now considering the putative Seiberg-dual pair with same number $N_f$ of flavors
and the ranks of gauge groups being $N$ and $N_f-N$, respectively, it is difficult to have
both theories in the pair to satisfy the bound (\ref{n=4-unitarity}).
Such cases are \cite{Kapustin:2010mh} $N_f=2N$ when $N_f$ is even, and $N_f=2N-1$
or $N_f=2N+1$ when $N_f$ is odd. The first case is self Seiberg-dual, and the next two
cases are Seiberg-dual to each other (with fixed $N_f$). So the only possible nontrivial
Seiberg duality with standard fixed point will be between the theory with $N_f=2N-1$ and
another with same $N_f$ and rank $N-1$.

However, even the last Seiberg duality has to be understood with care, because
there exists an operator which saturates the unitarity bound $\Delta\geq\frac{1}{2}$
when $N_f=2N-1$. The operator with scale dimension $\frac{1}{2}$ should correspond
to a free field, or a free twisted hypermultiplet \cite{Gaiotto:2008ak}. In particular,
the case with $N_f=N=1$ belongs to this case, in which case the naive Seiberg dual has rank
$N_f-N=0$ that the former cannot be dual to nothing. Thus, even the theory with $N_f=2N-1$
cannot be Seiberg-dual to its `naive dual' in the simplest sense. The modified proposal
is that the theory with $N_f\!=\!2N\!-\!1$ is dual to its naive dual times a decoupled theory
of a free twisted hypermultiplet \cite{Kapustin:2010mh}. As we shall see in detail,
the decoupled sector comes from the Abrikosov-Nielsen-Olesen (ANO) vortices
created by the monopole operator with dimension $\frac{1}{2}$.

Now let us study these dualities using vortex partition functions.

The Higgs vacua of the $\mathcal{N}=4$ theory form a hyper-K\"{a}hler moduli space.
On a subspace of this vacuum manifold with $\tilde{q}_i=0$, there exist BPS vortices
in the spectrum. This submanifold is compact and takes the form
of $\frac{U(N_f)}{U(N)\times U(N_f-N)}$. In particular, there is no moduli space if
$N_f=N$. The `naive' Seiberg dual pair have the same form of this moduli subspace.
In particular, we naturally identify the $U(N)\times U(N_f-N)$
global symmetries acting on the two moduli spaces.

We compare our vortex partition functions for the naive dual pairs, as functions of
$\mu_i$, $\mu_p$, $\gamma$, $\gamma^\prime$, $q$. In the quantum mechanical models
for the two types of vortices, the FI parameter $r$ appearing in section 2.2 corresponds
to the distance
between the two NS5-branes. Exchanging the two NS5-branes corresponds to changing
the sign of $r$. Thus, the vortex partition function for the `naive dual' theory can be
obtained from the original theory by tracing the effects of this sign change. In the
$\mathcal{N}=4$ theory, the only change is that the roles of $k\times N$ variable $q$
and the $(N_f-N)\times k$ variable $\tilde{q}$ are exchanged. So one is naturally led
to compare vortex/anti-vortex spectra in the dual pair as we explained in the previous
section, as the representations under $U(k)$ are conjugated after $q,\tilde{q}$ are
exchanged. Two vortex partition functions have different saddle points, either labeled by
division of $k$ into $N$ integers in the original theory, or into $N_f-N$ integers in the
naive dual. To obtain the partition function of the naive dual from the original one,
one should change the roles of $\mu_i$ and $\mu_p$, and further flip their signs.
The last sign flip is needed as the variables $q$/$\tilde{q}$
charged in $U(N)$ and $U(N_f-N)$ change their roles, making their charges
flip signs. This flip can be undone by flipping the signs of $\gamma,\gamma^\prime$,
as the index is manifestly invariant under the sign flips of
all $\mu_i,\mu_p,\gamma,\gamma^\prime$.

We first consider the index $I_N^{N_f}(q,\mu,\gamma,\gamma^\prime)$ with low values of $k$,
after expanding it as
\begin{equation}
  I_{N}^{N_f}=\sum_{k=0}^\infty q^kI_{N,k}^{N_f}(\mu,\gamma,\gamma^\prime)\ ,
\end{equation}
where $I_{N,0}^{N_f}\equiv 1$. At unit vorticity, $k=1$, we obtain (with $\tilde{R}=-R$)
\begin{equation}
  I^{N_f}_{N,1}=\frac{\sin(\gamma+\gamma')}{\sin2\gamma}
    \sum_{i=1}^N\prod_{j(\neq i)}^{N}\frac{\sinh\left(\frac{\mu_{ji}+2i(\gamma-\gamma')}{2}\right)}
    {\sinh\frac{\mu_{ji}}{2}}
    \!\!\prod_{p=N+1}^{N_f}\!\!\frac{\sinh\left(\frac{\mu_{pi}-2i(\gamma-\gamma')}{2}\right)}
    {\sinh\frac{\mu_{pi}}{2}}
\end{equation}
for the original partition function, and
\begin{equation}
  \tilde{I}^{N_f}_{N_f-N,1}=\frac{\sin(\gamma+\gamma')}{\sin2\gamma}\sum_{p=N+1}^{N_f}
  \prod_{j=1}^{N}\frac{\sinh\left(\frac{\mu_{jp}+2i(\gamma-\gamma')}{2}\right)}
    {\sinh\frac{\mu_{jp}}{2}}\prod_{q(\neq p)}^{N_f}
    \frac{\sinh\left(\frac{\mu_{qp}-2i(\gamma-\gamma')}{2}\right)}{\sinh\frac{\mu_{qp}}{2}}
\end{equation}
for the `dual' partition function. They apparently take very different forms,
as the first and second are sums over $N$ and $N_f-N$ terms, respectively. After summation,
we find that they are related in a simple manner. For simplicity, let us consider
the case in which $N_f\leq 2N$: the other case with $N_f\geq 2N$ can be obtained from
this by changing the roles of two theories. Then, one finds that
\begin{equation}\label{single-duality}
  I^{N_f}_{N,1}-\tilde{I}^{N_f}_{N_f-N,1}=\frac{\sin(\gamma+\gamma^\prime)}{\sin 2\gamma}
  \ \chi_{2N-N_f}(\gamma-\gamma^\prime)=I_{\rm com}(\gamma,\gamma^\prime)
  \chi_{2N-N_f}(\gamma-\gamma^\prime)\ ,
\end{equation}
where
\begin{equation}
  \chi_{2N-N_f}(\gamma-\gamma^\prime)=e^{i(2N-N_f-1)(\gamma-\gamma^\prime)}
  +e^{i(2N-N_f-3)(\gamma-\gamma^\prime)}+\cdots+e^{-i(2N-N_f-1)(\gamma-\gamma^\prime)}
\end{equation}
is the character for the $2N\!-\!N_f$ dimensional representation of $SU(2)$.
By definition, $\chi_0=0$. We have checked this expression for many cases, varying $N,N_f$.
Note that, even if $I^{N_f}_{N,1}$ and $\tilde{I}^{N_f}_{N_f-N,1}$ separately depend
on $\mu_i,\mu_p$, their difference on the right hand side does not. The result says that
the single vortex states in the `naive' dual pair are actually not the same. Rather,
the theory with larger gauge group rank $N$ ($> N_f-N$) has more states given by the
simple expression on the right hand side of (\ref{single-duality}). The right
hand side could be naturally explained if the excess states appear in a definite $SU(2)_L$
representation and are neutral in $J_R$ and $J_E$. In particular, when $N_f=2N-1$, the
above formula (\ref{single-duality}) becomes
\begin{equation}\label{single-duality-2}
  I^{2N-1}_{N,1}-\tilde{I}^{2N-1}_{N-1,1}=I_{\rm com}(\gamma,\gamma^\prime)\ ,
\end{equation}
implying that the excess state at $k=1$ is just one more single-particle state.
This can appear if the $U(N)$ theory is dual to the $U(N-1)$ theory (the naive dual)
times a decoupled twisted hypermultiplet with unit vorticity as suggested in \cite{Kapustin:2010mh}.

A more reassuring relation is found at $\mathcal{O}(q^2)$ order. We find that
\begin{equation}\label{two-duality}
  I^{N_f}_{N,2}-\tilde{I}^{N_f}_{N_f-N,2}=\frac{I_{2N-N_f}(\gamma,\gamma^\prime)^2
  +I_{2N-N_f}(2\gamma,2\gamma^\prime)}{2}+I_{2N-N_f}(\gamma,\gamma^\prime)\tilde{I}^{N_f}_{N_f-N,1}
\end{equation}
where
\begin{equation}\label{decoupled-single}
  I_{2N-N_f}(\gamma,\gamma^\prime)\equiv I_{\rm com}
  (\gamma,\gamma^\prime)\chi_{2N-N_f}(\gamma-\gamma^\prime)\ .
\end{equation}
Combined with the $k=1$ order results, this suggests that the exact
relation between the two vortex partition functions is
\begin{equation}\label{exact-duality}
  I^{N_f}_{N}(q,\mu,\gamma,\gamma^\prime)=\tilde{I}^{N_f}_{N_f-N}(q,\mu,\gamma,\gamma^\prime)
  \exp\left[\sum_{n=1}^{\infty}\frac{1}{n}I_{2N-N_f}(n\gamma,n\gamma^\prime)q^n\right]\ .
\end{equation}
Expanding both sides up to $\mathcal{O}(q^2)$, one recovers (\ref{single-duality}) and
(\ref{two-duality}). We have also checked (\ref{exact-duality}) at $\mathcal{O}(q^3)$ for
a few low values of $N_f,N$. At $N_f=2N-1$, the exponential factor on the right
hand side becomes
\begin{equation}
  \exp\left[\sum_{n=1}^\infty\frac{1}{n}I_{\rm com}(n\gamma,n\gamma^\prime)q^n\right]\ ,
\end{equation}
which is exactly the multi-particle index one obtains from a free twisted hypermultiplet
with unit vorticity: the single particle index is given by $I_{\rm com}(\gamma,\gamma^\prime)q$.
This precisely supports the Seiberg duality of \cite{Kapustin:2010mh}. It is also very natural
that this free field carries unit vorticity, as this decoupled sector is suggested from the
existence of a monopole operator with GNO charge $(1,0,0,\cdots,0)$ saturating the unitarity
bound. The above free field states should naturally be regarded as
being created by this monopole operator.

It is also interesting to find that the vortex partition functions of the two `naive'
dual pairs are related in a very simple manner for general $N\leq N_f$, although not
being completely equal. First of all, let us insert $N_f=N$ to
(\ref{exact-duality}). Then one obtains
\begin{equation}\label{local-index}
  I^N_N=\exp\left[\sum_{n=1}^{\infty}\frac{1}{n}I_N(n\gamma,n\gamma^\prime)q^n\right]
  =\exp\left[\sum_{n=1}^{\infty}\frac{1}{n}I_{\rm com}(n\gamma,n\gamma^\prime)\chi_N(n\gamma\!-\!n\gamma^\prime)q^n\right]\ ,
\end{equation}
as $\tilde{I}^N_0=1$ from the absence of vortices when the gauge group rank is zero.
Thus, when $N_f\!=\!N$, the index is independent of the flavor fugacities $\mu_i$ and
takes the form of the multiparticle states of unit vortices. From the single particle index
$I_{\rm com}\chi_N(\gamma-\gamma^\prime)q$ in the exponent, one finds $N$ different species
of ideal vortex particles. The above partition function may be implying
that the low energy theory could be a free theory of $N$ twisted hypermultiplets.
When $N=N_f=1$, one simply gets a free theory description of the (massless)
ANO vortex at low energy.

Secondly, inserting (\ref{local-index}) back to (\ref{exact-duality}), one obtains
\begin{equation}\label{exact-duality-2}
  I^{N_f}_{N}(q,\mu,\gamma,\gamma^\prime)=\tilde{I}^{N_f}_{N_f-N}(q,\mu,\gamma,\gamma^\prime)
  I^{2N-N_f}_{2N-N_f}(q,\gamma,\gamma^\prime)\ .
\end{equation}
One may interpret this as implying a novel form of IR duality in which
the theory with $N_f<2N$ is dual to the naive Seiberg dual times a decoupled sector,
given by the $U(2N\!-\!N_f)$ theory with $2N\!-\!N_f$ flavors. There should be very
nontrivial requirements for this to be true. Firstly, as the global flavor
symmetry of the naive duals matches to be $U(N)\times U(N_f-N)$ in the UV
description, the $U(2N\!-\!N_f)$ flavor symmetry of the latter decoupled factor is not visible.
So there should be a $U(2N\!-\!N_f)$ symmetry enhancement of the theory with $N_f<2N$ in IR,
for the above factorized duality to be true. At the level of vortex partition function,
the latter $U(2N-N_f)$ flavor symmetry is invisible due to the disappearance of its
chemical potentials in the partition function, as explained in the previous
paragraph. The way how such an IR $U(2N-N_f)$ symmetry enhancement could appear is
suggested by the vortex partition function itself. As the decoupled factor
on the right hand sides of (\ref{exact-duality-2}) and (\ref{exact-duality}) is a
multi-particle (or Plethystic) exponential of ideal vortex particles,
appearance of $2N\!-\!N_f$ species of decoupled vortex particles in IR could provide
$U(2N\!-\!N_f)$ enhanced symmetry which rotates them.

Such a generalized duality also makes sense if one considers massless sectors.
The Coulomb branches of both theories (at $\zeta\!=\!0$) have dimension $4N$,
precisely after including the decoupled sector to the naive Seiberg dual. 
By studying the Coulomb/Higgs moduli spaces, it was already noted in \cite{Gaiotto:2008ak} 
that theories with $N_f\!<\!2N$ have some free vector multiplets (or twisted hypermultiplets) 
in IR, as complete Higgsing is impossible. Our finding may be regarded as a concrete 
characterization of this observation as a generalized Seiberg duality.

One might think that the vortex partition function is a rather special quantity,
probing the $\tilde{q}_i=0$ region of the Higgs branch only. As a further support,
we also note that a factorization like (\ref{exact-duality-2}) was observed from the
3-sphere partition function, briefly mentioned in the conclusion of
\cite{Kapustin:2010mh}. Using various relations proved in \cite{Kapustin:2010mh},
one can easily show this factorization as follows. The 3-sphere partition function of
a supersymmetric gauge theory is a function of the FI parameter, which they call $\eta$,
and the real masses $m_i$ ($i=1,2,\cdots,N_f$). \cite{Kapustin:2010mh} obtains
\begin{equation}\label{3-sphere}
  Z_N^{N_f}(\eta,m_i)=\left(\begin{array}{c}N_f\\N\end{array}\right)
  \left(\frac{i^{N_f-1}e^{\pi\eta}}{1+(-1)^{N_f-1}e^{2\pi\eta}}\right)^N
  \left[e^{2\pi i\eta\sum_{j=1}^Nm_j}\prod_{j=1}^N\prod_{k=N+1}^{N_f}
  2\sinh\pi(m_j-m_k)\right]_{\{m\}}\ ,
\end{equation}
where $\left[\ \right]_{\{m\}}$ denotes symmetrization with the $N_f!$ permutations
on the $N_f$ mass parameters. The structure of the formula inside the parenthesis is
such that $N_f$ masses are divided into $N$ and $N_f-N$ groups.
Therefore, apart from the factor
\begin{equation}
  \left(\frac{i^{N_f-1}e^{\pi\eta}}{1+(-1)^{N_f-1}e^{2\pi\eta}}\right)^N
  e^{2\pi i\eta\sum_{j=1}^Nm_j}\ ,
\end{equation}
the expression is invariant under replacing $N$ by $N_f-N$, i.e. going to its
naive Seiberg-dual. In particular, one obtains
\begin{eqnarray}
  \hspace*{-1cm}Z_N^{N_f}(\eta,m_i)&=&\left(\begin{array}{c}N_f\\N\end{array}\right)
  \left(\frac{i^{N_f-1}e^{\pi\eta}}{1+(-1)^{N_f-1}e^{2\pi\eta}}\right)^N
  e^{2\pi i\eta\sum_{j=1}^{N_f}m_j}\left[e^{-2\pi i\eta\sum_{j=N+1}^{N_f}m_j}\prod_{j=1}^N\prod_{k=N+1}^{N_f}
  2\sinh\pi(m_j-m_k)\right]_{\{m\}}\nonumber\\
  &=&(-1)^{N(N_f-N)}\left(\frac{i^{N_f-1}e^{\pi\eta}}{1+(-1)^{N_f-1}e^{2\pi\eta}}
  \right)^{N}\left(\frac{i^{N_f-1}e^{-\pi\eta}}{1+(-1)^{N_f-1}e^{-2\pi\eta}}\right)^{N-N_f}
  e^{2\pi i\eta\sum_{j=1}^{N_f}m_j}Z^{N_f-N}_{N_f}(-\eta,m_i)\nonumber\\
  &=&\left(\frac{i^{N_f-1}e^{\pm\pi\eta}}{1+(-1)^{N_f-1}e^{\pm 2\pi\eta}}
  \right)^{2N-N_f}e^{2\pi i\eta\sum_{j=1}^{N_f}m_j}Z^{N_f-N}_{N_f}(-\eta,m_i)\nonumber\\
  &=&(-1)^{N_f(N_f-N)}\left(\frac{i^{(2N-N_f)-1}e^{\pm\pi\eta}}{1+(-1)^{(2N-N_f)-1}
  e^{\pm 2\pi\eta}}\right)^{2N-N_f}e^{2\pi i\eta\sum_{j=1}^{N_f}m_j}Z^{N_f-N}_{N_f}(-\eta,m_i)
  \nonumber\\
  &=&(-1)^{N_f(N_f-N)}Z_{2N-N_f}^{2N-N_f}(\pm\eta;\sum_{j=1}^{2N-N_f}M_j=
  \pm\sum_{j=1}^{N_f}m_j)Z_{N_f}^{N_f-N}(-\eta;m_i)\ .
\end{eqnarray}
Thus, apart from the possible $-1$ sign for odd $N_f(N_f-N)$, the partition function of
the theory with $N_f<2N$ factorizes into two, to the naive Seiberg-dual partition function
and another one with both $N,N_f$ replaced by $2N-N_f$.\footnote{The
extra $-1$ sign also exists for $N_f=2N-1$, omitted in \cite{Kapustin:2010mh}.
The $i^{N_f-1}$ factor in (\ref{3-sphere}) causes this sign.}

It should be interesting to study this possibility of novel IR fixed points further, and
hopefully to shed more light on possible phases of 3d supersymmetric theories. We hope the
clues provided by the vortex partition function in this paper and the 3-sphere partition
function of \cite{Kapustin:2010mh} could provide guiding information for uncovering some
aspects of this subject. Incidently, \cite{Willett:2011gp} studied the $\mathcal{N}=2$
Seiberg dualities of \cite{Aharony:1997gp} in the context of 3-sphere partition function
and Z-extremization, and made a similar observation that IR symmetry enhancement and
appearance of a free sector are needed. Also, studies of enhanced symmetry and novel IR
fixed points in 4 dimensions are made recently in \cite{Gaiotto:2012uq}, using the
superconformal indices.

\subsection{Aspects of $\mathcal{N}=3$ dualities from vortices}

Let us now consider the $\mathcal{N}=3$ (Yang-Mills) Chern-Simons-matter theories
with $U(N)_\kappa$ gauge group and $N_f$ fundamental hypermultiplets. These theories
have Seiberg duality as discussed in \cite{Giveon:2008zn}: the above theory is proposed
to be dual to the $U(N_f+|\kappa|-N)_{-\kappa}$ theory with $N_f$ hypermultiplets.
The duality is proposed to hold in the range $0\leq N\leq N_f+|\kappa|$. This duality has
been studied in quite a detail. The 3-sphere partition function was studied in
\cite{Kapustin:2010mh}, which proved mostly numerical agreements between the modulus
of the two Seiberg-dual partition functions. In \cite{Hwang:2011qt}, the superconformal
indices of some dual pairs are studied and agreements were shown for certain low values
of $N,\kappa,N_f$. In the discussion section, we shall point out a subtlety in this index
comparison for more general values of these parameters, and suggest a possible
resolution. Similar issues for $\mathcal{N}\!=\!2$ Seiberg dualities have been already
addressed in \cite{Bashkirov:2011vy}, which we also revisit later.

In this subsection, we study the proposed Seiberg-dual pair theories after deforming
them by an FI parameter, and also discuss the vortex partition function. As explained
in section 2, the FI deformed Chern-Simons-matter (or Yang-Mills-Chern-Simons-matter)
theory has many different branches of partially Higgsed vacua. The partially Higgsed
vacuum with unbroken $U(n)$ gauge symmetry should be dual to the vacuum with unbroken
$U(\kappa-n)$ symmetry \cite{Hanany:1996ie}. So to discuss Seiberg duality, one inevitably
has to understand the vortex spectrum in the (partially) unbroken phase.

As discussed in section 2, there exist two types of brane/string configurations
carrying nonzero vorticity. First type is the topological vortices given by the D1-brane
stretched between D3-branes corresponding to broken gauge groups and the flavor D3-branes
and/or the 5-brane, as shown in Fig \ref{n=3-brane}. Another possible type is the fundamental
string stretched between the D3-branes corresponding to the unbroken gauge symmetry and other
branes, also shown in Fig \ref{n=3-brane}. Since fundamental strings are charged under the
unbroken $U(n)$ or $U(\kappa-n)$ Chern-Simons gauge field, nonzero vorticity is
induced. This yields non-topological vortices \cite{Jackiw:1990pr,Eto:2010mu}.

Let us also consider their BPS masses. D1-brane vortices have masses which
are integer multiples of $2\pi\zeta$. The masses of fundamental strings are
integer multiples of $\frac{2\pi\zeta}{\kappa}$,
as this length is determined by a triangle formed by the $(1,\kappa)$ brane in
Fig \ref{n=3-brane}. So in general, when one compares the spetra of the
Seiberg-dual pair in partially broken phases with generic Chern-Simons level
$\kappa>1$, one would have both integral and fractional vortices.

Here we first comment on the
spectra when one of the pair theories is in the totally Higgsed phase.
In the brane picture, we have $n=0$ on the left side of Fig \ref{n=3-brane}.
Then, all vorticies in this theory are topological, having integer multiples
of $2\pi\zeta$ as their masses. On the other hand, the Seiberg-dual theory is
in a vacuum with unbroken $U(\kappa)_{-\kappa}$ Chern-Simons gauge symmetry.
So one might naively think that the dual theory would have fractional vortices
with massses being multiples of $\frac{2\pi\zeta}{\kappa}$, invalidating the
duality invariance of the spectrum. A possible resolution goes as follows.
The dynamics of $U(\kappa)_{\pm\kappa}$ Chern-Simons gauge fields, or the
Yang-Mills Chern-Simons gauge fields, with $\mathcal{N}=2,3$ supersymmetry is
supposed to be very nontrivial. In $U(n)_\kappa$ $\mathcal{N}=2,3$ YM-CS theory,
integrating out the fermions in the vector multiplet with mass $kg_{YM}^2$
at low energy yields a 1-loop shift to the $SU(n)$ part of the Chern-Simons level.
It shifts as $\kappa\rightarrow\kappa-n$ when $\kappa>0$, and oppositely
when $\kappa<0$ so that the absolute value of the level decreases. The point
$n=|\kappa|$ is special as the 1-loop corrected level vanishes. Thus, the $SU(\kappa)$
part of the theory is confining at low energy \cite{Witten:1999ds}. This is because the
remaining gauge dynamics is governed by pure $SU(\kappa)$ Yang-Mills theory at zero CS
level. As the BPS fundamental strings are in the fundamental representation of
$SU(\kappa)$, one should only consider those forming gauge singlets in the confining
phase. The only way of making gauge singlets with BPS matters
in fundamental representation is to form $SU(\kappa)$ baryons using totally
antisymmetric tensor. Thus, gauge singlet non-topological vortices
come in $\kappa$-multiples of the above fundamental string, with their
masses being multiples of $2\pi\zeta$.

Compared to the topological vortices, the classical
and quantum aspects of non-topological vortices seem to be relatively ill-understood.
So what we can do in generic case is predicting the quantum degeneracy of non-topological
vortices via duality by studying topological ones. However, in a simple case, we can do
more by using various effective treatments of non-topological vortices and compare with
dual topological vortices studied in this paper. The remaining part of
this section is devoted to this study.

Consider the theory with $N=N_f=1$ at CS level $\kappa$. (We shall
soon restrict to the case with $\kappa=1$ for detailed studies.) We consider the pure
Chern-Simons matter theory without Yang-Mills term. Turning off $\tilde{q}=0$ as before,
the classical bosonic equation of motion is derived from the following reduced
action\footnote{Normalization differs from that of \cite{Kim:2006ee}. Also, the gauge
fields there and here are related by $A_{\rm there}=-A_{\rm here}$, as the covariant derivative
there is different from ours, $D_\mu q=(\partial_\mu-iA_\mu)q$. Some of the equations and
solutions are also changed below, either due to this difference or just correcting typos there.}
\begin{equation}
  \mathcal{L}=\frac{\kappa}{4\pi}\epsilon^{\mu\nu\rho}A_\mu\partial_\nu A_\rho
  -|D_\mu q|^2-\frac{4\pi^2}{\kappa^2}|q|^2(|q|^2-\zeta)^2\ .
\end{equation}
The two minima $|q|=\zeta$ and $q=0$ of the potential correspond to the Higgs phase
and the symmetric phase. BPS equations for both topological/non-topological vortices
are given by
\begin{equation}
  (D_1\mp iD_2)q=0\ ,\ \ D_0q\mp\frac{2\pi i}{\kappa}q(|q|^2-\zeta)=0\ .
\end{equation}
In \cite{Kim:2006ee}, vortex domain wall was obtained
for $\kappa>0$ and upper signs of the BPS equations:
\begin{equation}\label{wall}
  q=(2\zeta)^{1/2}\sqrt{\frac{e^{2\pi x^1/\kappa}}{1+e^{2\pi x^1/\kappa}}}
  e^{-2\pi\zeta i(x^0+x^2)/\kappa}\ ,\ \ A_2=A_0=-\frac{\pi|q|^2}{\kappa}\ .
\end{equation}
This is a domain wall along the $x^2$ direction located at $x^1=0$, which
separates the symmetric phase $q=0$ in $x^1<0$ and the broken phase $q=\sqrt{\zeta}$
in $x^1>0$. The domain wall has the following linear vortex density and
monentum density along $x^2$ direction:
\begin{equation}
  \mathcal{B}=\int dx^1\ F_{12}=-\frac{2\pi\zeta}{\kappa} \ ,\ \
  \mathcal{P}=\int dx^1\ T_{01}=\frac{\pi\zeta^2}{\kappa}\ .
\end{equation}
Furthermore, as the BPS energy density is given solely by vorticity without
having domain wall tension, it was argued \cite{Kim:2006ee} that one can bend
this `tensionless domain wall' to yield more BPS solutions. The conjecture of
\cite{Kim:2006ee} is that, at least for large vorticity, the classical solution
for non-topological vortices can be approximated by a droplet of broken phase
$q\neq 0$ inside the symmetric phase with $q=0$, separated by a thin vortex domain
wall of arbitrary shape.\footnote{Of course one could think of bending the wall to
have the unbroken phase outside. Quantum mechanically, there should be a sense of
doing so. However, as this would yield a topological vortex with quantized classical
vorticity, we expect there to be a subtlety in the above argument at the classical level.}

It could be possible to quantize this system and count the degeneracy explicitly.
In this paper, leaving the full discussion of this problem as a future work,
we shall reproduce some characteristic aspects of non-topological vortices coming
from the tensionless domain wall picture, using the dual topological vortex
index. This would nontrivially support both Seiberg duality
as well as the tensionless domain wall picture for non-topological vortices.

As the vorticity and tangential linear monentum density is along the curve of
the domain wall, the charges of a closed-loop have the following behaviors. The
vorticity is proportional to the circumference $\ell$ of the boundary of
the broken phase region,
\begin{equation}
  k=-\frac{1}{2\pi}\oint\mathcal{B}=\frac{\zeta\ell}{\kappa}\ .
\end{equation}
On the other hand, for a closed loop the total momentum cancels
to zero while the angular momentum is proportional to
the area $A$ of the broken phase region:
\begin{equation}
  J=\oint \vec{x}\wedge\mathcal{P}d\vec{x}=-2\mathcal{P}A=-\frac{2\pi\zeta^2}{\kappa}A\ .
\end{equation}
We put a minus sign because $J$ is negative for non-topological vortices
with $q=0$ region outside the wall. This can be easily seen by noting that the unbroken
region is on the left side of the wall in (\ref{wall}), and bending the wall
to a non-topological vortex makes a clockwise circulation of the momentum $\mathcal{P}$
with $J<0$. As the circumference $\ell$ of a curve gives an upper bound for the area
$A$ of the region it surrounds by $\ell^2\geq 4\pi A$, the vorticity $k$
gives an upper bound to the angular momentum $J$ as
\begin{equation}\label{angular-bound}
  k^2\geq\frac{2}{\kappa}|J|\ .
\end{equation}
This is reminiscent of the angular momentum bounds for other familiar
2-charge systems. For instance, $\frac{1}{4}$-BPS 2-charge systems which can
be realized as wrapped D0-D4 or F1-momentum states all come with the angular
momentum bound $Q_1Q_2\geq|J|$, where $Q_1,Q_2$ are the two charges. As the electric
charge $Q$ of a non-topological vortex is $\kappa$ times the vorticity, the bound
(\ref{angular-bound}) may be written as $kQ\geq 2|J|$. Just from the viewpoint
of vortices, this upper bound on $J$ is not so obvious, as putting many vortices together
would naturally yield a bound on $J$ which is linear in $k$. It is really the collective
linear momentum $\mathcal{P}$ along the domain wall which creates much more angular
momentum than the naive expectation.

Below, we show this phenomenon at $\kappa\!=\!1$ by studying the dual topological vortices
in the Seiberg-dual theory. The case with $\kappa\!=\!1$ is much simpler as the dual
vacuum is in the totally broken phase, admitting topological vortices only.
The cases with $|\kappa|>1$ involve non-Abelian vortex dynamics in the
Seiberg-duals (whose domain wall description is not explored) and a mixture of
non-topological/topologica vortices in a given vacuum.

At $\kappa\!=\!1$, one has the $U(1)$ theory with $N_f\!=\!1$ hypermultiplet in the
unbroken phase. We take $n=1$ on the left side of Fig \ref{n=3-brane}. As the gauge
symmetry is unbroken, we only need to consider non-topological vortices discussed
above.\footnote{In Fig \ref{n=3-brane} with $N\!=\!n\!=\!1$, it may seem that there are
no D3-branes for the string to end on. It is clearer to move D5's to have it between 
NS5- and $(1,\kappa)$-branes \cite{Giveon:2008zn}. Then the string can have one end on 
the D5-brane.} In the
Seiberg-dual vacuum, on the right side of the figure, the $U(1)_{-1}$ gauge symmetry is broken
that it suffices for us to consider topological vortices. In this case, the duality
predicts the equality of the non-topological vortex spectrum on one side and the topological
vortex spectrum on the other. We shall study the partition function of the latter and
reproduce (\ref{angular-bound}) at $\kappa=1$ (up to a subtlety to be explained below).
The general formula (\ref{general-cs-index}) at $\kappa=1$ applies to the vortices with
$\int F_{12}<0$ at $r>0$ (i.e. on the left side of Fig \ref{n=3-brane}). To get the Seiberg
dual anti-vortices at $r<0$, one again exchanges $\mu_i$ and $\mu_p$ in the formula, and
then put extra minus signs for all $\mu$'s. More precisely, we are interested in the single
particle bound states. We numerically obtained the following single particle index
\begin{equation}
  I_{\rm sp}=\sum_{k=1}^\infty q^kI_{{\rm sp},k}(\mu,\gamma)\ ,\ \ I(q,\mu,\gamma)=\exp\left[
  \sum_{n=1}^\infty\frac{1}{n}I_{\rm sp}(q^n,n\mu,n\gamma)\right]
\end{equation}
till $\mathcal{O}(q^{23})$. For instance, the few leading terms are given by
\begin{eqnarray}\label{cs-single-index}
  I_{\rm sp}&=&qe^{\mu_1}e^{i\gamma}+q^2e^{2\mu_1}\frac{e^{5i\gamma}-e^{-i\gamma}}{1+e^{-4i\gamma}}
  +q^3e^{3\mu_1}\left(e^{13i\gamma}-e^{7i\gamma}\right)\\
  &&+q^4e^{4\mu_1}
  \left(e^{25i\gamma}-e^{19i\gamma}+e^{17i\gamma}-e^{15i\gamma}-e^{11i\gamma}
  +e^{9i\gamma}\right)\nonumber\\
  &&+q^5e^{5\mu_1}\left(e^{41i\gamma}-e^{35i\gamma}+e^{33i\gamma}-e^{31i\gamma}+e^{29i\gamma}
  -2e^{27i\gamma}+2e^{25i\gamma}+\cdots-e^{11i\gamma}\right)+\cdots\ .\nonumber
\end{eqnarray}
The maximal value of $-(J_R+2J_E)$ for given $k$ can be read off by identifying the term
with maximal power in $e^{i\gamma}$ at $\mathcal{O}(q^k)$. There are two cases that we
explain separately.

\begin{table}[t!]
$$
\begin{array}{c|cccccccccccc}
  \hline k&1&2&3&4&5&6&7&8&9&10&11&12\\
  \hline -2J_{\rm max}&1&&13&25&41&&85&113&145&&221&265\\
  \hline\hline k&13&14&15&16&17&18&19&20&21&22&23&\cdot\\
  \hline -2J_{\rm max}&313&&421&481&545&&685&761&841&925&1013&\\
  \hline
\end{array}
$$
\caption{Maximal values of the angular momentum $-(J_E+J_R/2)$}\label{maximal}
\end{table}
Firstly, when $k\neq 4p+2$ with an integer $p$, the terms with maximal angular momentum
all come with degeneracy $+1$, indicating that the shape of the domain wall curve is
indeed rigid so that no degeneracy is generated. The maximal values of
$-2J\equiv -2(J_R+2J_E)$ that we find in $z_{\rm sp}$ for
$k\leq 23$ are given in Table \ref{maximal}.
$-2J_{\rm max}$ denotes the maximal value of $-(2J_R+4J_E)$ that we find from the
single particle index, as the exponent multiplying $i\gamma$ as $e^{-i\gamma(2J_{\rm max})}$
in the index. Plotting $k$-$J_{\rm max}$, one easily finds
that we asymptotically find $J_{\rm max}\approx k^2$. As we do not expect the R-charge
$J_R$ to scale as a quadrature of $k$, we take it as the asymptotic growth of $2J_E$ and
find $|J_E|_{\rm max}\approx\frac{k^2}{2}$, confirming the property of non-topological
vortices. Moreover, it is easy to check the following
exact relation $-J_{\rm max}=2k^2-2k+1$ for $k\leq 23$ from Table \ref{maximal}.
This clearly shows that, ignoring the subdominant terms for large $k$, the upper bound
is quadratic in $k$ with the correct coefficient.

The single particle index for $k=4p+2$ is more complicated and actually hard to
understand from the effective domain wall description. The coefficients of $q^{4p+2}$
all take the form $\frac{e^{2i|J_{\rm max}|\gamma}+\cdots}{1+e^{-4i\gamma}}$, where
$J_{\rm max}=2k^2-2k+1$ takes the same form as other values of $k$. $\cdots$ denotes
a polynomial with smaller angular momenta. So apart from the $(1+e^{-4i\gamma})^{-1}$
factor, we find a similar upper bound in $J$. We currently do not understand the extra
factor at the moment. Perhaps the effective tensionless domain wall picture of
\cite{Kim:2006ee} might have a limitation at $k\neq 4p+2$ for some subtle unnoticed
reason. More study is needed to clearly understand this discrepancy. However, we still
find it amusing that $|J_E|_{\rm max}\approx\frac{k^2}{2}$ indeed holds in other cases.

\section{Discussions}

In this paper, we studied the supersymmetric partition function on
$\mathbb{R}^2\times S^1$ for topological vortices in 3d
$\mathcal{N}=4,3$ gauge theories, in
which a $U(N)$ gauge field is coupled to $N_f$ hypermultiplets.
The partition function admits a clear index interpretation for the local vortices
when $N_f=N$. The index interpretation is subtler for the semi-local vortices
for $N_f>N$, due to the non-compact internal zero modes. Even in the latter case,
the zero modes are lifted by the flavor `chemical potentials.' The partition
function is used to study 3d Seiberg dualities.

While studying these dualities, it becomes clear that the duality is exchanging
the light (or perhaps massless in the conformal point) vortices, just like
the 4d Seiberg duality exchanging elementary particles and magnetic monopoles, etc.
This emphasizes the importance of studying vortices and their partition functions
for a better understanding of Seiberg dualities, or more generally strongly coupled
IR physics, in 3d.

The vortex partition functions imply that there may be more possible Seiberg
dualities with $\mathcal{N}\!=\!4$ SUSY than those addressed in the literature.
Namely, the Seiberg dualities of UV theories with `standard IR fixed points' were
suggested and studied in \cite{Gaiotto:2008ak,Kapustin:2010mh} at $N_f=2N\mp 1$,
with a decoupled twisted hypermultiplet sector. As seen by the vortex partition
function (and also by the 3-sphere partition function as we reviewed), the
duality may extend to the whole window $0\leq N\leq N_f$, in which a UV theory
with $N_f<2N$ is suggested to be dual to the naive Seiberg dual times a decoupled
sector with $N,N_f$ replaced by $2N-N_f$. For this duality to hold, enhanced
IR symmetries and decoupled free sectors have to appear. It should be
interesting to study these issues further.

We also found interesting vortex spectrum in $\mathcal{N}=3$ Chern-Simons-matter
theories, but the structures of the vacua and vortex spectrum are much richer so that
more studies are required. We have compared the vortices in the theory with $N=N_f=1$
and $\kappa=\pm 1$, in which we found some nontrivial agreement between the
proposed Seiberg-dual pair.

There are several directions which we think are interesting.

It would first be interesting to have a definite index interpretation for
the partition function of semi-local vortices at $N_f\!>\!N$. In \cite{Collie:2009iz},
a parton-like interpretation for these vortex size moduli is given. More precisely,
they considered lump solitons in the $\mathbb{CP}^N$ sigma model, which are related in
IR to our vortices. Also, the partons from electrically charge particles in
\cite{Collie:2009iz} appear if we mirror dualize the theory we have been discussing
in this paper. A more challenging problem along this line would be the interpretation
of the index for 5d instantons \cite{Kim:2011mv}, perhaps with a similar partonic
picture which could shed light to the 6d $(2,0)$ SCFT in UV.

We would also like to see if the vortex partition function has any relation
to the superconformal index which counts magnetic monopole operators \cite{Kim:2009wb}.
This is conceptually well-motivated as monopole operators are basically vortex-creating
operators. Also, since the vortex partition function yields a good function of
chemical potentials in the conformal limit $\zeta\rightarrow 0$, it might be plausible
to seek for an alternative CFT interpretation of this quantity. The expression of the monopole
index in 3d SCFT is given in \cite{Kim:2009wb} as an infinite series expansion in the GNO
charges of monopoles. This contains infinitely many terms, which should be more efficiently
written in some cases. (See next paragraph for a related comment.) Trying to rewrite it
using the vortex partition functions could provide an alternative expression for the same
quantity. See \cite{Dimofte:2011py} for a related comment.

As a somewhat remotely related subject, we also remark on tests of 3d Seiberg dualities
with $\mathcal{N}=2,3$ supersymmetry in the literatures using monopole operators.
In particular, the $\mathcal{N}=3$ Seiberg dualities
of \cite{Giveon:2008zn} between Chern-Simons-matter theories are considered in detail.
Monopole operators in Chern-Simons-matter theories are more complicated than those
without Chern-Simons term, as magnetic fluxes induce nonzero electric charges which should
be screened by turning on matter fields. Spectrum of such monopoles has been studied
either by using localization technique \cite{Kim:2009wb} to calculate the index,
or by actually constructing semi-classical monopole solutions at large Chern-Simons level
\cite{Kim:2010ac}. Tests using the monopole index have been carried out in \cite{Hwang:2011qt}
for some low values of $N,N_f,\kappa$. However, if one considers the spectrum in full
generality for arbitrary $N,N_f,\kappa$, apparently one seems to find a problem about
R-charges of monopoles similar to the $\mathcal{N}\!=\!4$ monopoles of \cite{Gaiotto:2008ak}.
More concretely, the index measures the charge $R+2j_3$ with a chemical potential, where
$j_3$ is the angular momentum of operators on $\mathbb{R}^3$. This plays a role analogous
to the R-charge in the index. The lowest value of this charge for a given GNO charge
$H=(n_1,n_2,\cdots,n_N)$ can be obtained from the index, which is
\begin{equation}\label{R+2j}
  R+2j_3=\frac{N_f}{2}\sum_{i=1}^N|n_i|-\sum_{i<j}|n_i-n_j|+|\kappa|
  \sum_{i=1}^N\left(\frac{|n_i|}{2}+n_i^2\right)\ .
\end{equation}
Although the index only measures $R+2j_3$ charges, in $\mathcal{N}\!=\!3$
theories we can separately say what the values of $R$ and $j_3$ are. This is because they are
Cartans of $SU(2)_R$ and spatial $SO(3)$ rotations, which are both non-Abelian. As non-Abelian
charges are not renormalized along continuous deformation of the theory, one can trust the
values of $R$ and $j_3$ obtained from the deformed theory.
Similar calculation of non-Abelian R-charges was explained in \cite{Benna:2009xd}.
Using this property, one obtains
\begin{equation}\label{R-and-j}
  R=\frac{N_f+|\kappa|}{2}\sum_{i=1}^N|n_i|-\sum_{i<j}|n_i-n_j|\ ,\ \
  j_3=\frac{|\kappa|}{2}\sum_{i=1}^Nn_i^2\ .
\end{equation}
In particular, the expression for the R-charge as seen by the index takes the same
form as the R-charges (\ref{n=4-R-charge}) of $\mathcal{N}\!=\!4$ monopoles, after
replacing $N_f$ by $N_f+|\kappa|$. As the rank bound suggested for the $\mathcal{N}\!=\!3$
theory is $N\leq N_f+|\kappa|$, $N_f+|\kappa|$ plays the role of $N_f$ in many
places. In particular, if $N$ becomes close to $N_f+|\kappa|$,
one would have a similar problem of having R-charges, or even $R+2j_3$, becoming
too negative.

Practically, the expression for the index in \cite{Kim:2009wb} becomes of little
use in some cases. Introducing the fugacity $x$ for $R+2j_3$, the index is given as
an expansion with $x<1$ for given GNO charge. However, in various theories with $N$
close to $N_f+|\kappa|$, we find the following problem. The minimal value (\ref{R+2j}) of
$R+2j_3$ becomes negative for some GNO charge. Once we find a negative charge,
one can find more monopoles such that $R+2j_3$ is unbound from below. This implies that
an expansion in $x$ is ill-defined as one sums over all possible GNO charges. In fact,
terms with negative powers in $x$ should be forbidden for superconformal theories.
This problem arises only at the strongly coupled point in which
the 't Hooft coupling $N/k$ is not small.

The only way this pathological behavior can be eliminated from the index,
if we indeed have SCFT for all $N$ in the range $0\leq N\leq N_f+|\kappa|$,
seems to be that the above terms with negative powers in $x$ all cancel out with other
monopole contributions. Possible cancelations of some monopole contributions
to the index for $\mathcal{N}=2$ theories of \cite{Aharony:1997gp} were discussed in
\cite{Bashkirov:2011vy}. This problem emphasizes the need for a more
efficient expression for the index than those presented in \cite{Kim:2009wb},
perhaps using the vortex partition function. Also, seeking for a 3d analogue of
the recent study of the `diverging' 4d superconformal index \cite{Gaiotto:2012uq}
could be interesting.

Finally, it will be interesting to understand the vortex partition function of
the mass-deformed ABJM theory \cite{Hosomichi:2008jb} and learn more about the quantum
aspects of this system as well as its gravity dual. Some works in this direction have been
done in \cite{arXiv:1001.3153,arXiv:0905.1759,Lambert:2011eg}. In particular, the
Witten index for the vacua was calculated in \cite{arXiv:1001.3153}, both from QFT and
its gravity dual, fully agreeing with each other. However, the gauge/gravity duality of
the vortex spectrum poses a puzzle \cite{arXiv:0905.1759} at the moment.

\vskip 0.5cm

\hspace*{-0.8cm} {\bf\large Acknowledgements}
\vskip 0.2cm

\hspace*{-0.75cm} We are grateful to Tudor Dimofte, Dongmin Gang, Choonkyu Lee,
Kimyeong Lee, Sungjay Lee, Takuya Okuda, Jaemo Park, Masahito Yamazaki and Shuichi
Yokoyama for helpful discussions. This work is supported by the BK21 program of the
Ministry of Education, Science and Technology (JK, SK), the National Research Foundation
of Korea (NRF) Grants No. 2010-0007512 (HK, JK, SK) and 2005-0049409 through the Center
for Quantum Spacetime (CQUeST) of Sogang University (JK, SK, KL). SK would like to thank
the organizers of ``Mathematics and Applications of Branes in String and M-theory''
(Issac Newton Institute, Cambridge) and ``Classical and Quantum Integrable Systems''
(Bogoliubov Laboratory of Theoretical Physics, JINR, Dubna) for hospitality and support,
where part of this work was done.

\appendix

\section{SUSY and cohomological formulation}

In this appendix we construct the cohomological formulation of the vortex
quantum mechanics which is useful in the Witten index computation.
The quantum mechanical model for $\mathcal{N}=4$ vortices was introduced in section 2.2.
The lagrangian (\ref{QM-action}) preserves 4 real supersymmetries $Q_a$,
which are given by
\begin{eqnarray}
    Q_a A_t\!\!\!&\!\!=\!&\!\! i \bar\lambda_a \ , \quad \bar{Q}_a A_t = -i\lambda_a \nonumber \\
    Q_a \phi^I \!\!&\!\!=\!\!&\!\! i(\tau^I)_a^{\ b}\bar\lambda_b \ , \quad \bar{Q}_a\phi^I = i (\tau^I)_a^{\ b}\lambda_b \nonumber \\
    Q_a \lambda_b \!\!&\!\!=\!\!&\!\! (\tau^I)_{ab}\left(-D_t\phi^I
    +\frac{1}{2}\epsilon_{IJK}[\phi^J,\phi^K]\right)+i\epsilon_{ab} D \nonumber \\
    \bar{Q}_a\bar\lambda_b \!\!&\!\!=\!\!&\!\! (\tau^I)_{ab}\left(D_t \phi^I +\frac{1}{2}\epsilon_{IJK}[\phi^J,\phi^K]\right)-i\bar\epsilon_{ab} D
\end{eqnarray}
for the vector multiplet,
\begin{eqnarray}
    Q_a Z \!\!&\!\!=\!\!&\!\! \sqrt{2}\chi_a \ , \quad \bar{Q}_a Z^\dagger =- \sqrt{2}\bar\chi_a \nonumber \\
    \bar{Q}_a \chi_b \!\!&\!\!=\!\!&\!\! -i\sqrt{2}\epsilon_{ab} D_t Z-\sqrt{2}(\tau^I)_{ab} [\phi^I,Z] \nonumber \\
    Q_a \bar\chi_b \!\!&\!\! =\!\!&\!\! -i\sqrt{2}\epsilon_{ab} D_t Z^\dagger+\sqrt{2}(\tau^I)_{ab} [\phi^I,Z^\dagger]
\end{eqnarray}
for the adjoint chiral multiplet, and
\begin{eqnarray}
    Q_a q \!\!&\!\!=\!\!&\!\! \sqrt{2}\psi_{a} \ , \quad \bar{Q}_a q^\dagger =- \sqrt{2}\bar\psi_a  \nonumber \\
    \bar{Q}_a\psi_b \!\! &\!\!=\!\!&\!\! -i\sqrt{2}\epsilon_{ab} D_t q-\sqrt{2}(\tau^I)_{ab}\phi_Iq  \nonumber \\
    Q_a \bar\psi_b \!\!&\!\!=\!\!&\!\! -i\sqrt{2}\epsilon_{ab} D_t q^\dagger-\sqrt{2}(\tau^I)_{ab} q^\dagger\phi_I
\end{eqnarray}
for the $N$ fundamental chiral multiplets. Similarly, one can obtain the SUSY transformations of $N_f-N$
anti-fundamental chiral multiplets $\tilde{q},\tilde\psi$.
For $\mathcal{N}=3$ vorticies, the lagrangian differs by the term (\ref{CS-mechanics})
and only 2 real supercharges $Q_2$ (and its complex conjugation) are preserved.

To define the Witten index (\ref{index-trace}), we choose one supercharge among $Q_a$
\begin{eqnarray}
    Q\equiv \frac{1}{\sqrt{2}}(Q_2 + \bar{Q}_1) \ .
\end{eqnarray}
The index counts the BPS particles annihilated by $Q$. We can develop
a cohomological formulation using $Q$. Let us define
\begin{eqnarray}
    &&\phi \equiv A_t+\phi^3\ ,\quad \bar\phi \equiv -A_t+\phi^3
    \ , \quad \phi^\pm \equiv \frac{1}{\sqrt{2}}(\phi^1\pm i \phi^2) \nonumber \\
    &&\eta \equiv \sqrt{2}i(\lambda_1-\bar\lambda_2)\ , \quad \Psi \equiv \frac{i}{\sqrt{2}}(\lambda_1+\bar\lambda_2) \ .
\end{eqnarray}
The lagrangian and the $Q$ transformation can be rewritten with these new variables.
The $Q$ transformation for the vector multiplet is given by
\begin{eqnarray}\label{cohomology-vector}
    &&Q\phi =0\ , \quad Q\bar\phi = \eta\ , \quad Q\eta=[\phi,\bar\phi] \nonumber \\
    &&Q\phi^+ = i\lambda_2\ , \quad Q\phi^- = i\bar\lambda_1\ , \quad Q^2\phi^\pm = [\phi,\phi^\pm] \nonumber \\
    &&Q\Psi = \mathcal{E} \equiv -i[\phi_1,\phi_2]-D\ , \quad Q^2\Psi = [\phi,\Psi] \ ,
\end{eqnarray}
where we omitted time derivatives $\partial_t$ acting on the fields as we use a
`matrix model' like notation for convenience.
One can restore time derivatives by replacing $-iA_t$ by $D_t$ whenever we need.
We see that the square of the supercharge, $Q^2$, acting on the fields yields the gauge transformation generated by the complexified parameter $\phi$. Accordingly, $Q$ is nilpotent operator on-shell (we used the fermion equation of motion for the last equality in
(\ref{cohomology-vector})) up to the gauge rotation. This fact allows us to construct
$Q$ cohomology. Since off-shell nilpotency is required for localization, we introduce
an auxiliary scalar $H$ and modify the supersymmetry transformation as
\begin{eqnarray}
    Q\Psi = H\ , \quad QH = [\phi,\Psi] \ .
\end{eqnarray}
To have the off-shell invariant action the bosonic potential should also be changed as follows
\begin{eqnarray}
     -\frac{1}{2}{\rm tr}\left(\mathcal{E}^2\right) \quad \rightarrow \quad {\rm tr}\left(\frac{1}{2}H^2 -H\mathcal{E}\right) \ .
\end{eqnarray}
By integrating out $H$, we can recover the original action and the supersymmetry.

It is straightforward to generalize the cohomological formulation to the chiral multiplets.
We obtain the supersymmetry transformation
\begin{eqnarray}
    QZ = \chi_2\ , \quad Q\chi_2=[\phi,Z]
    \ , \quad Q\chi_1 = \mathcal{F}_Z \equiv -[\phi^1-i\phi^2,Z]\ , \quad Q^2\chi_1 =[\phi,\chi_1]\ ,
\end{eqnarray}
for the adjoint chiral multiplet,
\begin{eqnarray}
    &&Qq = \psi_2\ , \quad Q\psi_2 = \phi q \ ,\quad Q\psi_1 = \mathcal{F}_q \equiv -(\phi^1-i\phi^2)q \ , \quad Q^2 \psi_1 = \phi\psi_1 \nonumber \\
    &&Q\tilde{q} = \tilde\psi_2\ , \quad Q\tilde\psi_2 = -\tilde{q}\phi\ ,
    \quad Q\tilde\psi_1= \tilde{\mathcal{F}}_q \equiv \tilde{q}(\phi^1-i\phi^2)\ , \quad Q^2\tilde\psi_1 = -\tilde\psi_1\phi \ ,
\end{eqnarray}
for the $N$ fundamental and for the $N_f-N$ anti-fundamental chiral multiplets, respectively.
We again introduce auxiliary scalars $h_Z,h_q,\tilde{h}_q$ and find the following off-shell
supersymmetry
\begin{eqnarray}
    &&Q\chi_1 = h_Z\ , \quad Qh_Z =[\phi,\chi_1] \nonumber \\
    &&Q\psi_1 = h_q\ , \quad Qh_q = \phi\psi_1 \nonumber \\
    &&Q\tilde\psi_1 = \tilde{h}_q\ , \quad Q\tilde{h}_q = -\tilde\psi_1\phi\ .
\end{eqnarray}
The bosonic potential containing these auxiliary scalars is written as
\begin{eqnarray}
    {\rm tr}\left( h_Zh_Z^\dagger+h_qh_q^\dagger+\tilde{h}_q\tilde{h}_q^\dagger
    - (\mathcal{F}_Zh_Z^\dagger +\mathcal{F}_qh_q^\dagger +\tilde{\mathcal{F}}_q\tilde{h}_q^\dagger+c.c)\right)\ .
\end{eqnarray}
Collecting all the results, the bosonic part of the Euclidean lagrangian can be written as
\begin{eqnarray}
    L_B\!\!&\!\!=\!\!&\!\! {\rm tr}\left(\frac{1}{8}[\phi,\bar\phi]^2-\frac{1}{2}[\phi,\phi^I][\bar\phi,\phi^I]-\frac{1}{4}|[\phi-\bar\phi,Z]|^2 +\frac{1}{4}|[\phi+\bar\phi,Z]|^2-\frac{1}{2}H^2+H\mathcal{E} \right. \nonumber \\
    &&\quad \left.+ h_Zh_Z^\dagger+h_qh_q^\dagger+\tilde{h}_q\tilde{h}_q^\dagger
    - (\mathcal{F}_Zh_Z^\dagger +\mathcal{F}_qh_q^\dagger +\tilde{\mathcal{F}}_q\tilde{h}_q^\dagger+c.c)\right)
\end{eqnarray}
where $I=1,2$.

\section{Saddle points}
We evaluate the Witten index (\ref{index-trace}) using localization.
The index can be represented by a path integral, using the lagrangian
with Euclidean time $\tau$ ($t\equiv-i\tau$). The time $\tau$ is now periodic with
periodicity $\beta$ and the dynamical variables satisfy periodic boundary conditions
due to the insertion of $(-1)^F$ to the index,
which makes the path integral to be supersymmetric.
Indeed, the Hamiltonian of the Witten index is the square of the supercharge $Q^2=H$.
This implies that the  Witten index does not depend on the parameter $\beta$.

We also introduce chemical potentials $\mu_i,\gamma$ and $\gamma'$ to the path integral.
The boundary conditions of the fields are twisted by these chemical potentials.
There is an alternative way to deal with this twisting using the twisted time derivative.
Under the twisting the time derivative is shifted as
\begin{eqnarray}
    D_\tau \rightarrow D_\tau -\frac{\mu^i}{\beta}\Pi_i -i\frac{\gamma}{\beta}(2J)-i\frac{\gamma'}{\beta}(2J') \ .
\end{eqnarray}
Note that $J$ and $J'$ commute with the supercharge $Q$, so the deformed lagrangian is still invariant under $Q$. See \cite{Kim:2011mv} for details.

The index is independent of the continuous parameters $\beta,r$.
So we can take any convenient values of these parameters for the calculation.
We consider the limit $\beta\rightarrow 0,r \rightarrow \infty$, after which the
path integral is localized around the supersymmetric saddle points.
At the saddle point, all fermionic fields are set to zero and the bosonic fields are constrained by supersymmetry:
\begin{eqnarray}\label{saddle-equation}
   \!\!\! &&Q\eta = [\phi,\bar\phi]=0\ , \quad Q\psi_1 = \mathcal{F}_q = 0 \ , \quad Q\tilde\psi_1 = \tilde{\mathcal{F}}_q = 0 \ ,\quad Q\chi_1=\mathcal{F}_Z = 0 \nonumber \\
    \!\!\!&&Q\psi_2 = \phi q -q\frac{\mu}{\beta} +\frac{2i(\gamma J+\gamma'J')}{\beta}q = 0 \ , \quad Q\tilde\psi_2 = -\tilde{q}\phi+\frac{\mu}{\beta}\tilde{q} +\frac{2i(\gamma J+\gamma' J')}{\beta}\tilde{q} = 0  \nonumber \\
  \!\!\!  &&Q\chi_2 = [\phi, Z] +\frac{2i(\gamma J+\gamma'J')}{\beta}Z = 0  \ , \quad Q\Psi = -i[\phi^1,\phi^2]-\left([Z,Z^\dagger] + q q^\dagger -\tilde{q}^\dagger\tilde{q} -r\right)=0\ .\qquad \quad
\end{eqnarray}
We integrated out all the auxiliary scalars $H,h_Z,h_q,\tilde{h}_q$, and the
chemical potential $\mu$ here is a diagonal $N\times N$ matrix.
The last three equations on the first line can be solved by setting $\phi^1-i\phi^2 = 0$.
Using the $U(k)$ gauge transformation and $[\phi,\bar\phi]=0$ condition, we can take
the saddle point value of $\phi$ to be a diagonal $k\times k$ matrix.
$\bar\phi$ also becomes a diagonal $k\times k$ matrix at the saddle point,
but the exact value is not determined by the above equations.
It will be determined later by using the equation of motion of $\phi$.
The remaining equations reduce to
\begin{eqnarray}\label{SUSY-equation}
    &&\phi q^i -\frac{\mu_i-2i(\gamma+\gamma')R_q}{\beta}q^i=0\ , \quad \tilde{q}_p\phi-\frac{\mu_p+2i(\gamma+\gamma')R_{\tilde{q}}}{\beta}\tilde{q}_p=0 \ , \nonumber\\
    &&[\phi,Z]+\frac{4i\gamma}{\beta}Z=0 \ ,\quad  [Z,Z^\dagger]+q^iq_i^\dagger -\tilde{q}^{p\dagger} \tilde{q}_p = r \ .
\end{eqnarray}
These equations imply that the full solutions can be constructed by using the notion of $k$ dimensional vector space.
The $k\times k$ matrices $\phi,Z$ act as operators on the vector space and $q^i$ and
$\tilde{q}^{p\dagger}$ can be regarded as $N_f$ eigenvectors of $\phi$ with eigenvalues $\frac{\mu_i-2i(\gamma+\gamma')R}{\beta}$ and $\frac{\mu_p+2i(\gamma+\gamma')\tilde{R}}{\beta}$, respectively.
Two eigenvectors $q^i$ and $\tilde{q}^{p\dagger}$ are orthogonal to each other unless they are the same type,
namely $\tilde{q}_pq^i=0$, $q^\dagger_iq^j=0$ for $i\neq j$ and $\tilde{q}_p\tilde{q}^{r\dagger}=0$ for $p\neq r$.
Considering an eigenstate $|\lambda\rangle$ with $\phi|\lambda\rangle = \lambda |\lambda\rangle$ and acting the operator $Z$ on it,
one can obtain the other state with shifted eigenvalue, $Z|\lambda\rangle = |\lambda - \frac{4i\gamma}{\beta}\rangle$.
Therefore, $Z$ behaves as the raising operator shifting the eigenvalue of the states by $-\frac{4i\gamma}{\beta}$ and
its conjugate acts as the lowering operator shifting the eigenvalue in opposite way.
It is possible to obtain the complete basis of the $k$ dimensional vector space from the ground state, defined to be annihilated by $Z^\dagger$,
by acting $Z$ many times.

For $r>0$, we find $\tilde{q}=0$ from the last equation of (\ref{SUSY-equation}).
The same phenomenon happens to the instanton calculus and the proof is given in
\cite{Kim:2011mv}. A similar argument holds in our case and we can set $\tilde{q}$
to zero at the generic saddle point. Thus we only consider the eigenstates $q^i$
and their descendants
\begin{eqnarray}
    |m\rangle_i \propto Z^m q^i \ ,
\end{eqnarray}
with eigenvalues $\frac{\mu -2i(\gamma+\gamma')R -4im\gamma}{\beta}$ of $\phi$ for $m\ge 0$ and $i=1,2,\cdots,N$. As the vector space is finite dimensional, it will terminate at some number
$m$. There is a 1-to-1 correspondence between the set of these eigenstates and one dimensional
Young diagram. The number of the boxes in the Young diagram is determined by the number of
states in the corresponding set.
There are $N$ such Young diagrams and, since the vector space is $k$ dimensional, the total number of boxes in the Young diagrams should be $k$.
Therefore, the saddle points can be classified by the one dimensional $N$-colored Young diagrams with total box number $k$.
For the given colored Young diagrams, the explicit values of fields of the corresponding saddle point are determined by solving the last equation of (\ref{SUSY-equation}).

As an example, let us find the saddle point solutions for some low values of $k$, using
the above construction. We first consider the case with $k=1$. Here $q^i$ is simply
a number for each $i$. Only one of $N$ numbers, say $i$'th one, can be nonzero.
The vector $q^i$ has eigenvalue $\frac{\mu_i -2i(\gamma+\gamma)R}{\beta}$ of $\phi$.
It is annihilated by $Z$ as the total vector space is $k=1$ dimensional, so $Z=0$.
The last equation of (\ref{SUSY-equation}) fixes $q^i = \sqrt{r}e^{i\theta}$
where $\theta$ is the phase for the broken $U(1)$ on the $i$'th D3-brane.
The phase factor $\theta$ can be eliminated by the unbroken $U(1)$ gauge symmetry of the single vortex quantum mechanics.
We write the $i$'th saddle point as ${\tiny \yng(1)}_{\ i}$ from the colored Young diagram notation.
This can be understood as a single D1-brane bound to $i$'th D3-brane.

At $k=2$, there are two kinds of saddle points.
The first saddle point takes (${\tiny \yng(1)_{\ i},\yng(1)_{\ j}}$) form of the colored Young diagram.
This is a superposition of two $k=1$ saddle points where $q^i,q^j\, (i\neq j)$ contain nonzero components.
Here, $q^i$s are two dimensional vectors.
It corresponds to one D1-brane bound to $i$'th D3-brane and the other D1-brane bound to $j$'th D3-brane.
Using $U(2)$ gauge transformation we can write the solution as
\begin{eqnarray}
    q^i \!=\! \sqrt{r}(1\ 0) \ , \quad q^j \!=\! \sqrt{r}(0\ 1)\ , \quad
    \phi \!=\!{\rm diag}\left(\frac{\mu_i-2i(\gamma+\gamma')R}{\beta}\,,\, \frac{\mu_j-2i(\gamma+\gamma')R}{\beta}\right)\ ,\quad
\end{eqnarray}
with $Z=0$. This solves all equations in (\ref{SUSY-equation}).
We can also consider two phase factors $\theta_{1,2}$ for $q^i,q^j$ corresponding to the unbroken $U(1)^2$ gauge symmetry on two D3-branes,
but they can be eliminated by $U(1)^2\subset U(2)$ gauge transformation of two vortices.

The second saddle point is given by the colored Young diagram ${\tiny \yng(2)_{\ i}}$.
In this case, only one vector $q^i$ among $N$ vectors has nonzero component.
We can write it as $q^i = \lambda|1\rangle$ where $\phi|1\rangle = \frac{\mu_i-2i(\gamma+\gamma')R}{\beta}|1\rangle$.
We need one more state to form a two dimensional vector space. It will be obtained by acting $Z$ on $|1\rangle$ once.
Thus we find
\begin{eqnarray}
    |2\rangle \propto Z|1\rangle \ , \quad Z =c|2\rangle \langle1| \ ,
\end{eqnarray}
which implies that 2-1 component of $2\times 2$ matrix $Z$ gets nonzero value $c$.
The state $|2\rangle$ is killed by $Z$ so there is no more state.
Two eigenstates $|1\rangle,|2\rangle$ form
a complete basis of $k=2$ dimensional vector space.
The last equation of (\ref{SUSY-equation}) again fixes the undetermined constants $\lambda,c$ and, using the $U(2)$ gauge transformation, the solution is given by
\begin{eqnarray}
    q^i \!=\! \sqrt{2r}(1\ 0) \ , \quad \phi \!=\! {\rm diag}\left(\frac{\mu_i - 2i(\gamma+\gamma')R}{\beta},\frac{\mu_i -2i(\gamma+\gamma')R-4i\gamma}{\beta}\right)
    , \quad Z\!=\! \left(\begin{array}{cc}\!\!\! 0 & 0 \!\!\! \\ \!\!\! \sqrt{r} & 0 \!\!\!\end{array}\right). \quad
\end{eqnarray}
This solution illustrates two vortices bound to the single $i$'th D3-brane.

Finally, we explain the solution for $\bar\phi$.
The saddle point value of $\bar\phi$ is not fully determined by the equation (\ref{SUSY-equation}).
It only imposes the condition $[\phi,\bar\phi]=0$ which can be solved by taking a diagonal $\bar\phi$. We should use the equation of motion of $\phi$ (which is a Gauss' law constraint
for the $U(k)$ gauge symmetry) to determine the saddle point value of $\bar\phi$.
The variation $\delta\phi$ yields
\begin{eqnarray}\label{phibar}
    -\left[Z^\dagger,[\bar\phi,Z]-\frac{4i\gamma}{\beta}Z\right]+\frac{1}{2}\left(\bar\phi qq^\dagger
    +qq^\dagger\bar\phi\right)+q\frac{\mu-2i(\gamma+\gamma') R}{\beta}q^\dagger=0 \ ,
\end{eqnarray}
with $q$ and $Z$ taking the saddle point values. This equation is easily solved by setting $\bar\phi=-\phi$.
One can check it using the first three equations in (\ref{SUSY-equation}).

\section{Determinants}

We now compute the 1-loop determinant of the path integral around the saddle points obtained above.
We localize the path integral by taking the limit $\beta\rightarrow 0,r \rightarrow \infty$ since the Witten index is independent of these parameters.
Then the 1-loop determinant of the quadratic fluctuations with the classical action will give the exact result in this limit.
The quadratic terms of the bosonic fields around a generic saddle point is given by
\begin{eqnarray}\label{Lb}
    \!\!\!\!\!&& \!\!\!\!\!L_B^{(2)}=\frac{1}{8}\left(\delta\dot{\phi}+\delta\dot{\bar{\phi}}
    -[\delta{\phi},\bar{\phi}]-[\phi,\delta\bar{\phi}]\right)^2+|[\delta\phi_I,Z]|^2 + qq^\dagger\delta\phi_I\delta\phi_I + \frac{1}{4}|[\phi+\bar\phi,\delta{Z}] +[\delta\phi+\delta\bar\phi,Z]|^2\nonumber \\
    \!\!\!\!\!&& \!\!\!\!\!+\frac{1}{2}\left(\delta{\dot\phi}_I-[\phi,\delta \phi_I]
    -\frac{2i(\gamma J+\gamma'J')}{\beta}\delta\phi_I\right)\left(\delta{\dot\phi}_I
    +[\bar\phi,\delta \phi_I]-\frac{2i(\gamma J+\gamma'J')}{\beta}\delta\phi_I\right) \nonumber \\
    \!\!\!\!\!&& \!\!\!\!\! +\left(\delta \dot{Z}^\dagger+\frac{1}{2}[\delta Z^\dagger,\phi-\bar\phi]
    +\frac{1}{2}[Z^\dagger,\delta\phi-\delta\bar\phi]+\frac{4i\gamma}{\beta}\delta Z^\dagger\right)
    \left(\delta \dot{Z}-\frac{1}{2}[\phi-\bar\phi,\delta Z]
    -\frac{1}{2}[\delta\phi-\delta\bar\phi,Z]-\frac{4i\gamma}{\beta}\delta Z\right)\nonumber \\
    \!\!\!\!\!&& \!\!\!\!\! +\left(\delta\dot{q}^{\dagger}+\frac{1}{2}\delta q^{\dagger}(\phi-\bar\phi)
    +\frac{1}{2}q^{\dagger}(\delta\phi-\delta\bar\phi)-\frac{\mu}{\beta}\delta q^{\dagger}
    +\frac{2i(\gamma+\gamma') R_q}{\beta}\delta q^{\dagger}\right)\nonumber \\
    \!\!\!\!\! && \!\!\!\!\! \quad \times\left(\delta\dot q -\frac{1}{2}(\phi-\bar\phi)\delta q
     -\frac{1}{2}(\delta\phi-\delta\bar\phi)q  +\delta q\frac{\mu}{\beta}-\frac{2i(\gamma+\gamma')R_q}{\beta}\delta q_i\right) \nonumber \\
    \!\!\!\!\!&& \!\!\!\!\! +\left(\delta\dot{\tilde{q}}^{\dagger}-\frac{1}{2}(\phi-\bar\phi)\delta\tilde{q}^{\dagger}
    +\delta\tilde{q}^{\dagger}\frac{\mu}{\beta}+\frac{2i(\gamma+\gamma') R_{\tilde{q}}}{\beta}\delta\tilde{q}^{\dagger}\right)
    \left(\delta\dot{\tilde{q}}+\frac{1}{2}\delta\tilde{q}(\phi-\bar\phi)-\frac{\mu}{\beta}\delta\tilde{q}
    -\frac{2i(\gamma+\gamma') R_{\tilde{q}}}{\beta}\delta\tilde{q}\right)\nonumber \\
    \!\!\!\!\! && \!\!\!\!\!+\frac{1}{4}|(\phi+\bar\phi)\delta q+(\delta\phi+\delta\bar\phi)q|^2
     +\frac{1}{4}|\delta\tilde{q}(\phi+\bar\phi)|^2
     +\frac{1}{2}\left(\delta q q^{\dagger} +q \delta q^{\dagger} +[\delta Z,Z^\dagger]+[Z,\delta Z^\dagger]\right)^2\ ,
\end{eqnarray}
where $I=1,2$ and we used the facts $\phi_I=0,\tilde{q}=0$ at the saddle points.
Note that all the coefficient are quadratures of $\frac{\gamma}{\beta},\frac{\gamma'}{\beta},\partial_\tau$
and the saddle point values of the fields $\phi,Z,q$.
Here $\partial_t\sim \frac{1}{\beta}$ because the time direction is compactified with the radius $\beta$.
Also, the bosonic fields take the saddle point values proportional to $\frac{1}{\beta}$ or $\sqrt{r}$.
Therefore, when $\beta\rightarrow0,r \rightarrow \infty$, the above quadratic terms dominates other higher order terms so that the saddle point approximation can be applied.
Similar argument reduces the fermionic action to the quadratic action around the saddle points.
So we can obtain the exact value of the Witten index of the vortex moduli space by
evaluating the 1-loop integral of the bosonic terms (\ref{Lb}) and the fermionic quadratic terms given by
\begin{eqnarray}
    \!\!\!\!\!&&\!\!\!\!\!L_F^{(2)} = -\bar\lambda_1\left(\dot\lambda^1 +[\bar\phi,\lambda^1]
    -\frac{2i(\gamma-\gamma')}{\beta}\lambda^1\right)-\bar\lambda_2\left(\dot\lambda^2
    -[\phi,\lambda^2]\right) \nonumber \\
    \!\!\!\!\!&&\!\!\!\!\!-\bar\chi_1 \left(\dot\chi^1+[\bar\phi,\chi^1]
    -\frac{4i\gamma}{\beta}\chi^1\right)
    -\bar\chi_2 \left(\dot\chi^2-[\phi,\chi^2] -\frac{2i(\gamma+\gamma')}{\beta}\chi^2\right) \nonumber \\
    \!\!\!\!\!&&\!\!\!\!\!-\bar\psi_1\left(\dot\psi^1 +\bar\phi\psi^1+\psi^1\frac{\mu}{\beta}-\frac{2i(\gamma+\gamma')R_q}{\beta}\psi^1\right)
    -\bar\psi_2\left(\dot\psi^2 -\phi\psi^2+\psi^2\frac{\mu}{\beta}
    -\frac{2i(\gamma+\gamma')R_q-2i(\gamma-\gamma')}{\beta}\psi^2\right) \nonumber \\
    \!\!\!\!\!&&\!\!\!\!\!-\bar{\tilde{\psi}}_1\left(\dot{\tilde{\psi}}^1-\tilde{\psi}^1\bar\phi
    -\frac{\mu}{\beta}\tilde{\psi}^1-\frac{2i(\gamma+\gamma') R_{\tilde{q}}}{\beta}\tilde\psi^1\right)
    -\bar{\tilde{\psi}}_2\left(\dot{\tilde{\psi}}^2+\tilde{\psi}^2\phi-\frac{\mu}{\beta}\tilde{\psi}^2
    -\frac{2i(\gamma+\gamma')R_{\tilde{q}}-2i(\gamma-\gamma')}{\beta}\tilde\psi^2\right)\nonumber \\
    \!\!\!\!\!&&\!\!\!\!\!+\sqrt{2}i\left(\bar\chi^a[\bar\lambda_a,Z]+[Z^{\dagger },\lambda^a]\chi_a
    +q\bar\psi^a\bar\lambda_a+\lambda^a\psi_aq^{\dagger}\right)
\end{eqnarray}

We will provide the detailed computation of the determinant for one vortex.
For $k=2$, we have also performed this calculation which also confirms the general
result of section 2.

For $k=1$, we can ignore the commutators of the adjoint field, and we can also set $Z$ to zero
from the saddle point analysis, which makes the calculation much easier.
We first consider the bosonic part that gives the following contribution to the index.
\begin{enumerate}
\item $\delta Z$ : The quadratic action of the scalar $Z$ is given by
	\begin{eqnarray}
		\delta Z^\dagger\left(\frac{2\pi in}{\beta} +\frac{4i\gamma}{\beta}\right)\left(-\frac{2\pi i n}{\beta} -\frac{4i\gamma}{\beta}\right)\delta Z\ , \nonumber
	\end{eqnarray}
	for $n$'th Fourier mode of $Z$ where $Z\sim e^{-\frac{2\pi in}{\beta}\tau}$. The determinant is given by
	\begin{eqnarray}
		\left[\mathcal{N}^2\sin^22\gamma\right]^{-1} \ ,
	\end{eqnarray}
	where $\mathcal{N} \equiv -\frac{2i}{\beta^{1/2}}\prod_{n\neq 0}\left(\frac{-2\pi in}{\beta^{1/2}}\right)$.

\item $\delta\phi^{1,2}$ : The action is given by
	\begin{eqnarray}
		\delta\phi_{-}\left[\left(\frac{2\pi in}{\beta}-2i\frac{\gamma-\gamma'}{\beta}\right)\left(-\frac{2\pi in}{\beta}+2i\frac{\gamma-\gamma'}{\beta}\right)-r\right]\delta\phi_{+} \ ,\nonumber
	\end{eqnarray}
	where $\phi_{\pm}\sim \phi^1\pm i\phi^2$. The 1-loop contribution is
	\begin{eqnarray}
		\left[\mathcal{N}^2\sin\left(\gamma-\gamma' -i\sqrt{\frac{r\beta^2}{2}}\right)\sin\left(\gamma-\gamma' +i\sqrt{\frac{r\beta^2}{2}}\right)\right]^{-1} \ .
	\end{eqnarray}
\item $\delta{\tilde{q}}_p$ : The action is given by
    \begin{eqnarray}
        -\delta \tilde{q}^{\dagger p} \left(\frac{2\pi in}{\beta}-\frac{\mu_{i}-\mu_p-2i(\gamma+\gamma')(R_q+R_{\tilde{q}})}{\beta}\right)^2\delta\tilde{q}_p \ ,\nonumber
    \end{eqnarray}
    whose one loop determinant is
    \begin{eqnarray}
        \prod_{p}\left[\mathcal{N}^2\sinh^2\left(\frac{\mu_i-\mu_p-2i(\gamma+\gamma')(R_q+R_{\tilde{q}})}{2}\right)\right]^{-1} \ .
    \end{eqnarray}
\item $\delta q^{j\neq i}$ : The action is given by
	\begin{eqnarray}
		\delta q_{\dagger j} \left(\frac{2\pi in}{\beta}+\frac{\mu_i-\mu_j}{\beta}\right)\left(-\frac{2\pi in}{\beta}-\frac{\mu_i-\mu_j}{\beta}\right)\delta q^j \ ,\nonumber
	\end{eqnarray}
	whose determinant is
	\begin{eqnarray}
		\prod_{j\neq i}\left[\mathcal{N}^2 \sinh^2\left(\frac{\mu_i-\mu_j}{2}\right)\right]^{-1} \ .
	\end{eqnarray}

\item $\delta\phi,\delta\bar\phi, q_i$ : The fluctuation of $q^i$ is taken to be
    \begin{eqnarray}
        q^i = e^{i\theta}\left(\sqrt{r} + \frac{\delta\rho}{\sqrt{2}}\right)\ .\nonumber
    \end{eqnarray}
    The action is given by
    \begin{eqnarray}
        \frac{1}{2}(\delta\dot{\rho})^2 + r (\delta\rho)^2 +r\left(\dot\theta - \delta A_\tau\right)^2 +\frac{1}{2}(\delta\dot\phi^3)^2 + r (\delta\phi^3)^2\ .\nonumber
    \end{eqnarray}
    We choose the gauge $\theta=0$, then the Faddeev-Popov determinant is simply 1. We will compute the integral
    \begin{eqnarray}
        \int [\sqrt{2r}d\rho]d(\delta A_\tau)d(\delta\phi^3)]{\rm exp}\left[-\int d\tau\left(
        \frac{1}{2}(\delta\dot{\rho})^2 + r (\delta\rho)^2 +r\left(\delta A_\tau\right)^2 +\frac{1}{2}(\delta\dot\phi^3)^2 + r (\delta\phi^3)^2\right)\right]\ , \nonumber
    \end{eqnarray}
    and it gives the result
    \begin{eqnarray}
        \left[\mathcal{N}^2\sinh\sqrt{\frac{r\beta^2}{2}}\right]^{-1}\ .
    \end{eqnarray}
\end{enumerate}
We can also compute the fermionic determinants from the fermion quadratic terms.
The results are as follows.
\begin{enumerate}
\item $\chi_a$ :
	The determinant is given by
	\begin{eqnarray}
		\mathcal{N}^2\sin2\gamma\,\sin(\gamma+\gamma')\ .
	\end{eqnarray}
\item $\tilde\psi_p^a$ : The determinant is given by
    \begin{eqnarray}
        \!\!\!\!\!\!\!\!\!\!\mathcal{N}^2\prod_p\sinh\left(\!\frac{\mu_i-\mu_p-2i(\gamma+\gamma')(R_q+R_{\tilde{q}})}{2}\!\right)
        \sinh\left(\!\frac{\mu_i-\mu_p-2i(\gamma+\gamma')(R_q+R_{\tilde{q}})+2i(\gamma-\gamma')}{2}\!\right).\quad
    \end{eqnarray}

\item $\psi_{j}^a$ with $j\neq i$ : The determinant becomes
	\begin{eqnarray}
		\mathcal{N}^2\prod_{j\neq i}\sinh\left(\frac{\mu_j-\mu_i}{2}\right)\sinh\left(\frac{\mu_j-\mu_i+2i(\gamma-\gamma')}{2}\right)\ .
	\end{eqnarray}

\item $\bar{\lambda}^2,\psi_{i1}$ : The action is given by
	\begin{eqnarray}
		\big(\lambda_2\;\; \bar\psi^{i1}\big)\left(\begin{array}{cc}-\frac{2\pi in}{\beta}& i\sqrt{2r} \\ -i\sqrt{2r} & -\frac{2\pi in}{\beta}\end{array}\right)\left(\begin{array}{c}\bar\lambda^2 \\ \psi_{i1}\end{array}\right)\ , \nonumber
	\end{eqnarray}
	and the determinant is given by
	\begin{eqnarray}
		\mathcal{N}^2\sinh^2\sqrt{\frac{r\beta^2}{2}}\ .
	\end{eqnarray}

\item $\bar\lambda^1,\psi_{i2}$ : The action is given by
	\begin{eqnarray}
		\big(\lambda_1\;\;\bar\psi^{i2}\big)\left(\begin{array}{cc}-\frac{2\pi in}{\beta}-2i\frac{\gamma-\gamma'}{\beta} & -i\sqrt{2r} \\ i\sqrt{2r} & -\frac{2\pi in}{\beta}+2i\frac{\gamma-\gamma'}{\beta}\end{array}\right)\left(\begin{array}{cc}\bar\lambda^1 \\ \psi_{i2}\end{array}\right)\ , \nonumber
	\end{eqnarray}
	The determinant is given by
	\begin{eqnarray}
		\mathcal{N}^2 \sin\left(\gamma-\gamma'-i\sqrt{\frac{r\beta^2}{2}}\right)\sin\left(\gamma-\gamma'+i\sqrt{\frac{r\beta^2}{2}}\right) \ .
	\end{eqnarray}
\end{enumerate}
Collecting all the determinants, we can calculate the index corresponding to the saddle point labeled by ${\tiny \yng(1)_{\ i}}$.
One can see the cancellation of terms including the parameters $\beta,r$ between bosonic and fermionic contributions. The remaining quantity does not depend on those parameters as we
expected. Summing over all $N$ different saddle points the full index in one vortex sector
is given by
\begin{eqnarray}
    I_{k=1} \!=\! \frac{\sin(\gamma+\gamma')}{\sin2\gamma}
    \sum_{i=1}^{N}\prod_{j\neq i}^{N}\frac{\sinh\frac{\mu_{ji}+2i(\gamma-\gamma')}{2}}{\sinh\frac{\mu_{ji}}{2}}
    \!\!\prod_{p=N+1}^{N_f}\!\!\frac{\sinh\frac{\mu_{pi}+2i(\gamma+\gamma')(R+\tilde{R})
    -2i(\gamma-\gamma')}{2}}
    {\sinh\frac{\mu_{pi}+2i(\gamma+\gamma')(R+\tilde{R})}{2}}\ ,
\end{eqnarray}
where $\mu_{ij}=\mu_i-\mu_j$.

\end{document}